\documentclass[amsmath,amssymb,aps,floatfix,nofootinbib,longbibliography]{revtex4-2}
\pdfoutput=1
\usepackage{graphicx}
\usepackage{appendix}
\usepackage{amsfonts}
\usepackage{comment}
\usepackage[bookmarks=false,colorlinks]{hyperref}
\hypersetup{urlcolor=blue, citecolor=blue}
\usepackage[utf8x]{inputenc} 
\usepackage[normalem]{ulem}

\usepackage{ragged2e}

\usepackage{multirow}

\def \bit{\begin{itemize}}
\def \eit{\end{itemize}}
\def \beq{\begin{equation}}
\def \eeq{\end{equation}}
\def \bea{\begin{eqnarray}}
\def \eea{\end{eqnarray}}
\def \[{\left[}
\def \]{\right]}
\def \({\left(}
\def \){\right)}
\def \lb{\left\{}
\def \rb{\right\}}
\def \lp{\left|}
\def \rp{\right|}
\def \l.{\left.}
\def \r.{\right.}
\def \<{\left\langle}
\def \>{\right\rangle}
\def \ga{\gamma}

\def \g5{{\gamma^5}}
\def \al{{\alpha}}
\def \be{{\beta}}
\def \si{\sigma}
\def \ep{\epsilon}
\def \vep{\varepsilon}

\def \bc{{\bar c}}

\def \nn{\nonumber}
\def \nl{\nn\\}

\def \cA{{\cal A}}
\def \bcA{{\cal \bar{A}}}
\def \cB{{\cal B}}
\def \cM{{\cal M}}
\def \cL{{\cal L}}
\def \cO{{\cal O}}

\def \Re{{\rm Re}}
\def \Im{{\rm Im}}

\def \s{\sqrt{2}}

\def \dAFB{{\Delta A_{FB}}}
\def \Dst{{D^*}}
\def \bclX{{b \to c \ell X}}
\def \bctaunu{{b \to c \tau \nu_\tau}}
\def \hh{{\hat{h}}}
\def \hL{{\hat{L}}}
\def \hl{{\hat{\ell}}}
\def \mDs{m_{D^*}}
\def \dhh{{\delta\hat{h}}}
\newcommand{\zcb}{z_{cb}}
\newcommand{\wcb}{w_{cb}}
\def \as{{\alpha_s}}
\def \oB{\overline{B}}

\allowdisplaybreaks

\begin{document}
\title{Implications for the $\Delta A_{FB}$ anomaly in
${\bar B}^0\to D^{*+}\ell^- {\bar\nu}$ using
a new Monte Carlo Event Generator }

%%%%%%%%%%%%%%%%%%%%% Author Names and Affiliations here %% %%%%%%%%%%%%%%%%%%%%%%%%%%%%%%%%%%%%%%%%%%%%%%%%%%%%%%%%%%%

\author{Bhubanjyoti Bhattacharya}
\email{bbhattach@ltu.edu}
\affiliation{Department of Natural Sciences, Lawrence Technological University, Southfield, MI 48075, USA}

\author{Thomas E. Browder}
\email{teb@physics.hawaii.edu}
\affiliation{Department of Physics and Astronomy, University of Hawaii, Honolulu, HI 96822, USA}

\author{Quinn Campagna}
\email{qcampagn@go.olemiss.edu}
\affiliation{Department of Physics and Astronomy, \\
108 Lewis Hall, University of Mississippi, Oxford, MS 38677-1848, USA}

\author{Alakabha Datta}
\email{datta@phy.olemiss.edu}
\affiliation{Department of Physics and Astronomy, \\
108 Lewis Hall, University of Mississippi, Oxford, MS 38677-1848, USA}

\author{Shawn Dubey}
\email{sdubey@hawaii.edu}
\affiliation{Department of Physics and Astronomy, University of Hawaii, Honolulu, HI 96822, USA}

\author{Lopamudra Mukherjee}
\email{lmukherj@olemiss.edu }
\affiliation{Department of Physics and Astronomy, \\
108 Lewis Hall, University of Mississippi, Oxford, MS 38677-1848, USA}

\author{Alexei Sibidanov}
\email{sibid@hawaii.edu}
\affiliation{Department of Physics and Astronomy, University of Hawaii, Honolulu, HI 96822, USA}

%%%%%%%%%%%%%%%%%%%%%%%%%%%%%%%%%%%%%%%%%%%%%%%%%%%%%%%%%%

\begin{abstract}
Recent experimental results in $B$ physics from Belle, BaBar and LHCb suggest new physics (NP) in the weak $b\to c$ charged-current processes. Here we focus specifically on the decay modes $\overline{B}^0\to D^{*+}\ell^- \bar{\nu}$ with $\ell = e$ and $\mu$. The world averages of the ratios $R_D$ and $R_D^{*}$ currently differ from the Standard Model (SM) predictions by $3.4\sigma$ while recently a new anomaly has been observed
in the forward-backward asymmetry measurement, $A_{FB}$, in $ \overline{B}^0\to D^{*+}\mu^- \bar{\nu}$ decay. It is found that
$\Delta A_{FB} = A_{FB}(B\to D^{*} \mu\nu) - A_{FB} (B\to D^{*} e \nu)$ is around $4.1\sigma$ away from the SM prediction in an analysis of 2019 Belle data.
In this work we explore possible solutions to the $\Delta A_{FB}$ anomaly and point out correlated NP signals in other angular observables. These correlations between angular observables must be present in the case of beyond the Standard Model physics.
We stress the importance of $\Delta$ type observables that are obtained by taking the difference of the observable for the muon and the electron mode. These quantities cancel form factor uncertainties in the SM and allow for clean tests of NP. 
These intriguing results also suggest an urgent need for improved simulation and analysis techniques in $\overline{B}^0\to D^{*+}\ell^- \bar{\nu}$ decays. Here we also describe a new Monte Carlo Event-generator tool
based on EVTGEN that we developed to allow simulation of the NP signatures in $\overline{B}^0\to D^{*+}\ell^- \nu$, which arise due to the interference between the SM and NP amplitudes. 
We then discuss prospects for improved observables sensitive to NP couplings with 1, 5, 50, and 250 ab$^{-1}$ of Belle II data, which seem to be ideally suited for this class of measurements.
\end{abstract}

\maketitle

\section{Introduction}

A powerful way to study physics beyond the Standard Model (SM) is via
virtual effects of new particles, not present in the SM, in low energy experiments. These virtual effects can in many cases  probe mass scales beyond the reach of present or proposed colliders, where the new particles are expected to appear. There is also the possibility that beyond the Standard Model physics comes in the form of weakly coupled light new states. These new states are more likely to be detected at low energy, high precision experiments. In this work we will focus on charged current semileptonic $B$ decays,
$\oB^0\to D^{*+}\ell^- \bar{\nu}$ with $\ell = e$ and $\mu$.
These decays originate from the underlying quark-level transitions $b \to c \ell^- \bar{\nu}_\ell$, where $\ell = e, \mu,$ or $\tau$. At the hadron level they manifest as decays such as ${\bar B} \to D^{(*)} \ell^- \bar{\nu}_\ell$.

The charged-current decays $B\to D^{(*)} \tau \nu_\tau$ have been measured by the BaBar, Belle and LHCb experiments. Discrepancies with SM predictions of $R_{D^{(*)}}^{\tau \ell} \equiv \cB(\oB \to D^{(*)} \tau^{-} {\bar\nu}_\tau)/\cB(\oB \to D^{(*)} \ell^{-}{\bar\nu}_\ell)$ ($\ell = e,\mu$) \cite{BaBar:2012obs, BaBar:2013mob, LHCb:2015gmp, Belle:2015qfa, Belle:2016ure, Belle:2016dyj, LHCb:2017smo, Belle:2017ilt, LHCb:2017rln, Belle:2019gij} have been observed thus far. The SM predictions and the corresponding World-Averaged experimental results from the Heavy Flavor Averaging Group (HFLAV) \cite{HFLAV:2019otj} are shown in Table \ref{tab:obs_meas}. The  deviation from the SM in $R_D^{\tau \ell}$ and $R_{D^*}^{\tau \ell}$ (combined) has a significance of $3.4\sigma$ %\cite{Watanabe:2017mip}. 
\cite{HFLAV:2019otj}.
These measurements suggest the presence of NP that is lepton-flavor universality violating (LFUV) in $\bctaunu$ decays. 
\begin{table}[htb]
\begin{tabular}{|c|c|c|} \hline
Observable & SM Prediction & Measurement (WA) \\
\hline
$R_{D^*}^{\tau/\ell}$ & $0.258 \pm 0.005$ \cite{HFLAV:2019otj} & $0.295 \pm 0.011 \pm 0.008$ \cite{HFLAV:2019otj} \\
$R_{D}^{\tau/\ell}$ & $0.299 \pm 0.003$  \cite{HFLAV:2019otj} & $0.340 \pm 0.027 \pm 0.013$ \cite{HFLAV:2019otj} \\
$R_{D^*}^{\mu/e}$ & $\sim 1.0$ & $1.04 \pm 0.05 \pm 0.01$ \cite{Belle:2017rcc} \\ \hline
\end{tabular}
\caption{Measured values of observables that suggest NP in $\bctaunu$. Measurements presented in this table refer to World Averages (WA). Note that in \cite{Biswas:2021pic}, the most recent lattice data from \cite{FermilabLattice:2021cdg} on $B \to D^* \ell \nu$ form factors were used to obtain the SM prediction for $R_{D^*}^{\tau/\ell}$, $0.2586 \pm 0.0030$.}
\label{tab:obs_meas}
\end{table}
%%%%%%%%%%%%%	

We will focus on the decay $\oB^0 \to D^{*+} \ell^- \bar{\nu}$ as a laboratory to explore NP effects in $b \to c \ell^- \bar{\nu}_\ell$ transitions. At leading order, the $\oB^0\to D^{*+}\ell^-{\bar\nu}$ transitions proceed via the SM. However, new interactions can affect these decays. In experiment, the underlying transition is $\bclX$ where the invisible state $X$ can be a left-handed (LH) neutrino (part of the SM LH doublet of leptons) or a light right-handed (RH) singlet neutrino. Here we will focus on NP scenarios that produce only LH neutrinos in the final state.

 Although theoretical work on NP has concentrated on the semi-leptonic $\tau$ modes, where experimental statistics are limited, attention is now also being paid to the semi-leptonic muon and electron modes where large data samples will be available. For example, scaling the Belle results in \cite{Belle:2018ezy} to Belle II at 50 ab$^{-1}$ we expect a yield of $8\times 10^6$ events in each of the muon and electron modes. Similarly, scaling the BaBar results in \cite{BaBar:2007ddh} on $B \to D^* \ell \nu$ with a fully reconstructed hadronic tag, we expect $3\times10^5$ events with no background. 

An additional advantage is that the missing neutrino momentum can be calculated from kinematic constraints of $e^+ e^-$ production at the $\Upsilon(4S)$ and the angular distributions can be fully reconstructed. Unlike the $\tau$, which is detected through its decay products, the muon and electron are directly detected in experiment. In contrast, for semi-leptonic $B$ decays to the $\tau$ lepton, the final state contains one or more additional neutrinos from the $\tau$ decay, which complicates the situation. Examining NP in the muon mode is further motivated by the anomalous $(g-2)_\mu$ measurements \cite{Muong-2:2021ojo} as well as by the neutral-current LFUV $B$ anomalies in the $b \to s \mu^+\mu^-$ decays (see for example, Ref.~\cite{LHCb:2021trn}). At first glance, when studying the $B$ anomalies within the framework of an Effective Field Theory (EFT), these anomalies may appear unrelated. However, within an SMEFT framework
NP in the $b\to s\mu^+\mu^-$ transition could imply NP in the $b\to c \mu^-{\bar\nu}_\mu$ decay \cite{Bhattacharya:2014wla}. In this article, therefore, we will focus on the muon and electron modes, assuming that the electron decay mode is well described by the SM, but NP contributions are allowed in the muon mode.

\begin{figure}[t]
\includegraphics[width=0.7\textwidth]{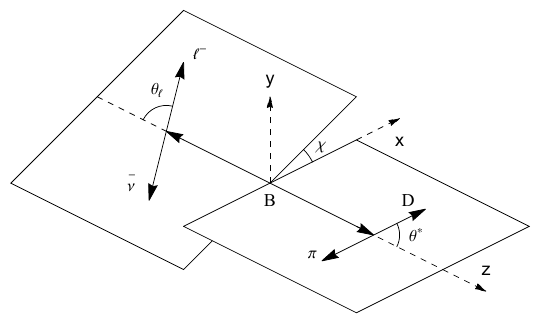}
\caption{Schematic diagram defining various angles in $\oB\to D^*(\to D\pi)\ell^-{\bar\nu}$ decay \cite{Bhattacharya:2019olg}. We have aligned the coordinate axes so that the decaying $\oB$ meson is at rest at the origin and in this frame the momentum of the $D^*$ meson is oriented along the z-axis. Subsequent decays are shown in the rest frames of the corresponding object that is decaying -- $D^* \to D\pi$ is in the rest frame of the $D^*$ and a virtual particle decays into $\ell^-{\bar\nu}$. The polar angles, $\theta^*$ and $\theta_\ell$, are respectively defined in these subsequent rest frames, while the azimuthal angle, $\chi$, is defined in the rest frame of the $\oB$ meson.}
    \label{fig:bdstarplanes}
\end{figure}

Although hints for NP have appeared in the ratio of rates such as $R_{D^{(*)}}$, establishing NP and diagnosing the type of NP will require examination of deviations from the SM in other observables as well. Several observables can be constructed
from a complete differential distribution of events using helicity angles. Fig.~\ref{fig:bdstarplanes} shows a schematic definition of the three helicity angles in $\oB \to D^{*} (\to D\pi)\ell^-{\bar\nu}$.

Angular observables are even more interesting as these may provide one or more unambiguous signals for NP. One such sensitive angular observable is the forward-backward asymmetry of the charged lepton, $A_{FB}$, which can be reconstructed as the difference between the number of leptons with the lepton's helicity angle, $\theta_\ell$ (see Fig.~\ref{fig:bdstarplanes}), greater and less than $\pi/2$. Another observable is $S_3$, which can be reconstructed as an asymmetric integral over the angle $\chi$, which measures the difference between the decay planes of the $D^*$ and the lepton-neutrino system (see Fig.~\ref{fig:bdstarplanes}). There are additional interesting and correlated angular observables, such as $S_5$ and $S_7$, which require asymmetric integrals over multiple helicity angles. In Ref.~\cite{Bhattacharya:2019olg}, it was shown that NP in the $\mu$ modes can also be detected in the CP-violating triple-product terms, like $S_7$, in the angular distribution \cite{Duraisamy:2013pia,Duraisamy:2014sna}. Some previous work in the literature on the effects of new physics in angular observables of semileptonic B-meson decays can be found in \cite{Hurth:2020rzx, Rajeev:2020aut, Ciuchini:2020gvn, Hurth:2020ehu, Becirevic:2022tsj, Becirevic:2022bev}.

A non-zero $A_{FB}$ is present in both the muon and electron channels in the SM due to interference between different helicity amplitudes of the virtual $W$ boson. However, in a $\Delta$-type observable\footnote{Such observables were first proposed in Ref.~\cite{Capdevila:2016ivx} for angular analyses study on $B \to K^* \ell \ell$ decay}, $\Delta A_{FB} = A_{FB}^\mu - A_{FB}^e$, where one considers the difference between the muon and electron channels, the SM contributions approximately cancel, except for a small residual effect due to the dependence on the muon mass close to its threshold. Furthermore, we find that the observable $\Delta A_{FB}$ has reduced sensitivity to hadronic uncertainties in form factors. %
Therefore, any deviation from the SM prediction for $\Delta A_{FB}$ is likely due to NP effects. Recently, using the tables of Belle data from Ref.~\cite{Belle:2018ezy}, an anomaly in $\Delta A_{FB}$ was reported in Ref.~\cite{Bobeth:2021lya}. This could be a signature of 
LFUV NP \cite{Bobeth:2021lya,Carvunis:2021dss,Datta:2022czw}.

LFUV NP in the electron and muon sectors is tightly constrained by the measurement of the ratio of rates $R_{D^{(*)}}^{\mu e} \equiv \cB(\oB \to D^{(*)} \mu^{-} {\bar\nu}_\mu)/\cB(\oB \to D^{(*)} e^{-} {\bar\nu}_e)$ which is $1.04 \pm 0.05$ \cite{Belle:2017rcc}. We restrict ourselves to NP Scenarios in which a deviation of at most 3\% from unity is allowed, which could be tested in the future. Even if the effects of LFUV NP are small in the ratios of decay rates, larger effects may be visible in the angular distributions.

In this paper, we discuss various solutions to explain the $\Delta A_{FB}$ anomaly.
The framework we use is based on a Monte Carlo generator to simulate a realistic experimental environment. Hence, in this work, we describe a newly developed Monte Carlo (MC) Event-generator tool \cite{Bhattacharya:2022cna} to allow simulations of the NP signatures in $B\to D^*\ell \nu$ arising due to the interference between SM and NP amplitudes. We employ our MC tool primarily to study semi-leptonic decays with a muon and electron in the final state. We assume that the electron decay mode is well described by the SM, but allow for NP contributions in the muon mode. Using this MC tool we generate results for three distinct scenarios with different NP couplings that are consistent with current data and can explain the $\Delta A_{FB}$ anomaly, while remaining consistent with other constraints. Furthermore, using MC simulations we demonstrate that $\Delta$-type observables, such as $\Delta A_{FB}$ and $\Delta S_5$, eliminate most QCD uncertainties from form factors and allow for clean measurements of NP. We introduce correlated observables that improve the sensitivity to NP. We also discuss prospects for improved observables sensitive to NP couplings with the expected 50 ab$^{-1}$ of Belle II data, which seems to be ideally suited for this class of measurements. These measurements may also be possible at LHCb and other hadron collider experiments. We provide both integrated observables, for the benefit of current experimental analyses, and distributions of the observables as a function of $q^2$. We also suggest experimental requirements on $q^2$ and on laboratory lepton momenta to optimize sensitivity to NP and reduce systematics.

The layout of the remainder of this article is as follows. In Section \ref{sec:th}, we discuss the theoretical basis of the full angular distribution for ${\bar B\to D^*\ell^-{\bar\nu}}$ in an effective theory framework. In Sections \ref{sec:npimpl}, \ref{sec:npsig}, and \ref{sec:npsens}, we present the implementation of our NP MC tool, the signatures of and sensitivity to NP respectively. 
%Section \ref{sec:future} describes improvements to be implemented in the future and 
finally we conclude in Section \ref{sec:conc}.

\section{Theory} \label{sec:th}

In the study of NP in charged-current semi-leptonic $B$ decays it is useful to adopt an EFT framework. In an EFT description of the $b\to c\ell^-{\bar\nu}$ decays, one writes down all possible dimension-six four-quark operators at the scale of the $b$-quark mass. The effective Hamiltonian that describes SM and NP effects can be expressed as,
\bea
{\cal H}_{\rm eff} &=& \frac{G_F V_{cb}}{\sqrt{2}} \Bigl\{
\left[(1 + g_L)\,{\bar c}\gamma_\al (1 - \gamma_5) b + g_R \, {\bar c} \gamma_\al (1 + \gamma_5) b \right]
{\bar \mu} \gamma^\al (1 - \gamma_5) \nu_\mu \nn\\
&& \hskip15truemm
+~\left[ g_S \, {\bar c} b + g_P \, {\bar c} \gamma_5 b \right] {\bar \mu} (1 - \gamma_5) \nu_\mu
+ g_T \, {\bar c} \sigma^{\al\be} (1 - \gamma_5) b
{\bar \mu} \sigma_{\al\be} (1 - \gamma_5) \nu_\mu\Bigr\} + h.c.~,
\label{4fermi_NP}
\eea
where the factors $g_X$, $X = L, R, S, P,$ and $T$, are coupling constants that describe NP effects. As indicated earlier, we have only included LH neutrinos in this EFT, however, we have allowed for both LH and RH NP couplings.

Based on the effective Hamiltonian of Eq.~(\ref{4fermi_NP}), one can express the decay amplitude for the process $\oB\to D^*(\to D\pi)\ell{\bar\nu}$ as \cite{Bhattacharya:2020lfm,Bhattacharya:2019olg},
\bea
\cM &=& \frac{4\,G_F V_{cb}}{\s} \Bigg\{\<D\pi\lp\bc\ga^\mu\[(1 + g_L)P_L + g_R P_R\] b\rp\oB\>
({\bar\ell}\ga_{\mu}P_L\nu) ~~ \nl
&&\hspace{15truemm} +~\<D\pi\lp\bc\(g_{S_L}P_L + g_{S_R} P_R\) b\rp\oB\>({\bar\ell}P_L\nu) + g_T \<D\pi\lp\bc\si^{\mu\nu} b\rp\oB\>({\bar\ell}\si_{\mu\nu}P_L\nu)\Bigg\} ,~~ \label{eq:ol}
\eea
where $P_{R,L} = (1 \pm \gamma_5)/2$. This decay amplitude contains several hadronic matrix elements that describe the $\oB\to D^* \to D\pi$ transitions through LH and RH scalar and vector currents, as well as a tensor current. The $D^*\to D\pi$ decay is mediated solely by the strong force, so that
\bea
\<D\pi|D^*(k,\ep)\> &=& \epsilon\cdot(p_D - p_\pi),
\eea
where $p_{D(\pi)}$ is the four-momentum of the $D(\pi)$, $k = p_D + p_\pi$ is the four-momentum of the $D^*$ and $\epsilon$ is its polarization. Note that these satisfy the on-shell condition $k\cdot\epsilon = 0$. 

The remaining parts of the hadronic matrix elements that appear in Eq.~(\ref{eq:ol}) are (see, for example, \cite{Sakaki:2013bfa}) :
\bea
\<D^*(k,\ep)\lp\bc\ga_\mu b\rp\oB(p)\> &=& -{\it i}\vep_{\mu\nu\rho\si}\ep^{*\nu}p^\rho k^\si\frac{2V(q^2)}{m_B + m_{D^*}} ,~~ \\
\<D^*(k,\ep)\lp\bc\ga_\mu\ga^5 b\rp\oB(p)\> &=& \ep^*_{\mu}(m_B + m_{D^*})A_1(q^2) - (p + k)_\mu(\ep^*\cdot q)\frac{A_2(q^2)}{m_B + m_{D^*}} ~~ \nl
&& \hspace{2truecm} -~ q_\mu(\ep^*\cdot q)\frac{2m_{D^*}}{q^2}[A_3(q^2) - A_0(q^2)] ,~~ \\
\<D^*(k,\ep)\lp\bc\ga^5 b\rp\oB(p)\> &=& -(\ep^*\cdot q)\frac{2 m_{D^*}}{m_b + m_c} A_0(q^2) ,~~ \\
\<D^*(k,\ep)\lp\bc\si_{\mu\nu} b\rp\oB(p)\> &=& \vep_{\mu\nu\rho\si}\lb-\ep^{\rho*}(p+k)^\si T_1(q^2) + \ep^{\rho*}q^\si\frac{m_B^2 - m_{D^*}^2}{q^2}[T_1(q^2) - T_2(q^2)]\r. ~~ \nl
&&\hspace{1truecm} \l. +~ 2\frac{\ep^*\cdot q}{q^2}p^\rho k^\si\[T_1(q^2)-T_2(q^2) - \frac{q^2}{m^2_B - m^2_{D^*}}T_3(q^2)\]\rb ~~
\label{eq:hadronic-matrix-elem}
\eea
where $p$ is the four-momentum of the $B$ meson, $q$ represents the four-momentum of the lepton-neutrino pair, while $m_{B(D^*)}$ represents the mass of the $B (D^*)$ meson. Here, $V, A_0, A_1, A_2, A_3, T_1, T_2$ and $T_3$ are the relevant form factors for a $\oB \to V$ transition. The BGL \cite{Boyd:1997kz}, CLN \cite{Caprini:1997mu} and HQET \cite{Bordone:2019vic} parameterizations for these form factors  are given in Appendix \ref{sec:ffs}. For the Levi-Civita tensor, $\varepsilon_{\mu\nu\rho\sigma}$, we use the convention $\varepsilon_{0123} = +1$. 

For easy comparison with similar literature in the field, below we present an alternative notation and its connection to the notation used in this article. Following the presentation in Ref.~\cite{Bobeth:2021lya}, the effective Lagrangian that describes $b\to c\ell^-{\bar\nu}$ transitions can be written as
\bea
\cL &=& -~\frac{4G_F}{\s}\sum\limits_i C_i{\cO}_i + h.c.~, \label{eq:bobeth}
\eea
where $i = V_L, V_R, S_L, S_R,$ and $T$, and $C_i$ represents the Wilson Coefficient (WC) corresponding to the operator $\cO_i$. Note the negative sign added to this Lagrangian in order to obtain the correct sign for the SM term (see for example Eq.~(20.90) in \cite{Peskin:1995ev} with errata in \cite{Peskin-errata}). The WC's can be easily converted into the NP coupling constants that appear in Eq. (\ref{4fermi_NP}) as follows.
\beq
C_{V_L} ~=~ 1 + g_L~,~~\quad
C_{V_R} ~=~ g_R~,~~\quad
C_{S_L} ~=~ g_S - g_P~,~~\quad
C_{S_R} ~=~ g_S + g_P~,~~\quad
C_{T} ~=~ g_T~.~~
\eeq
Note that only $C_{V_L}$ has both SM and NP parts while all other WCs are NP only. Furthermore, for a $\oB\to V$ transition, where $V$ denotes a vector meson, the scalar matrix element $\langle V|\bar{q}b|B\rangle = 0$. As a consequence, the following condition must be imposed,
\bea
C_{S_R} + C_{S_L} ~=~ 2\,g_S ~=~ 0~.~~ \label{eq:nuconstraint}
\eea
Thus, there are only four independent NP parameters that can be used to describe the decay $\oB\to D^* \ell^-{\bar\nu}$ process, namely $g_L, g_R, g_P$, and $g_T$. We will use the $g_i$ parameters to describe the results and plots presented in this article.

One can now express the differential decay distribution for $\oB\to D^* (\to D\pi)\ell^-{\bar\nu}$ as a function of four kinematic variables -- $q^2$ and three helicity angles $\theta^*, \theta_\ell,$ and $\chi$ (see Fig.~\ref{fig:bdstarplanes} for a schematic diagram defining these angles) -- in the following form. 
\bea
\frac{d^4\Gamma}{dq^2\,d\cos\theta^*\,d\cos\theta_\ell\,d\chi} &=& \frac{9}{32\pi}\[\(I_1^s\sin^2\theta^* + I_1^c\cos^2\theta^*\) + \(I_2^s\sin^2\theta^* + I_2^c\cos^2\theta^*\)\cos2\theta_\ell\r. \nl
&& \hspace{1truecm}+~I_3\sin^2\theta^*\sin^2\theta_\ell\cos2\chi + I_4\sin2\theta^*\sin2\theta_\ell\cos\chi + I_5\sin2\theta^*\sin\theta_\ell\cos\chi \nl
&& \hspace{1truecm}+~\(I_6^c\cos^2\theta^* + I_6^s\sin^2\theta^*\)\cos\theta_\ell + I_7\sin2\theta^*\sin\theta_\ell\sin\chi \nl
&& \hspace{1truecm} \l.+~I_8\sin2\theta^*\sin2\theta_\ell\sin\chi + 
I_9\sin^2\theta^*\sin^2\theta_\ell\sin2\chi\], \label{eq:angdist}
\eea
where the 12 coefficients $I_i^{(s,c)}(q^2)$ (i = 1,\ldots,9) can be expressed in terms of eight helicity amplitudes that in turn depend on the NP parameters $g_L, g_R, g_P$, and $g_T$. For brevity, the exact dependence of the coefficient functions, $I_i^{(s,c)}$ is given in Appendix \ref{sec:iiscs}. The distribution for the CP-conjugate process is obtained with the following transformation, $\theta_l \to \pi- \theta_l$ and $\chi \to \pi+\chi$. The various helicity amplitudes transform as
$\cA_{SP}  \to  -\bcA_{SP},~\cA_t \to -\bcA_t,~\cA_{0(,T)} \to \bcA_{0(,T)},~\cA_{||(,T)} \to \bcA_{||(,T)}, ~\cA_{\perp(,T)} \to -\bcA_{\perp(,T)} (\cA_{\pm(,T)} \to \bcA_{\mp(,T)})$ leading to the angular coefficients transformations $I_{1,2,3,4,7}^{(a)} \to \bar{I}_{1,2,3,4,7}^{(a)}$ and $I_{5,6,8,9}^{(a)} \to -\bar{I}_{5,6,8,9}^{(a)}$\footnote{Our convention is similar to the LHCb convention for the $B^0 \to K^{*0} \ell^+ \ell^-$ decay where $\theta_\ell$ is defined as the angle between $K^{*0} (\bar{K}^{*0})$ and $\mu^+ (\mu^-)$ for the $B^0 (\oB^0)$ decay leading to the transformations $I_{1,2,3,4,5,6}^{(a)} \to \bar{I}_{1,2,3,4,5,6}^{(a)}$ and $I_{7,8,9}\to -\bar{I}_{7,8,9}$ for CP conjugation with $\chi \to 2\pi - \chi$ \cite{Gratrex:2015hna}. Alternatively, when $\theta_\ell$ is defined as the angle between $K^{*0} (\bar{K}^{*0})$ and the lepton $\ell^-$ for the $B^0 (\oB^0)$ decay while $\chi$ is the angle between the $K^\pm \pi^\mp$ and the $\ell^+ \ell^-$ planes in both cases, the angular coefficients transform as $I_{1,2,3,4,7}^{(a)} \to \bar{I}_{1,2,3,4,7}^{(a)}$ and $I_{5,6,8,9}^{(a)} \to -\bar{I}_{5,6,8,9}^{(a)}$ for the CP conjugate process with $\theta_\ell \to \theta_\ell -\pi$ and $\chi \to -\chi$ \cite{Altmannshofer:2008dz, Bobeth:2008ij}. Note that, in all of these conventions, including ours, the $\frac{d^4(\Gamma + \bar{\Gamma})}{dq^2 d\cos \theta^* d\cos \theta_\ell d\chi}$ distribution for the untagged decay retains the contribution from the ``true'' CP violating terms \cite{Datta:2003mj, Gronau:2011cf}.}.
Note that if one writes ${\cal{A}}= |A| e^{ i \phi + i \delta}$, then ${\cal{\bar{A}}}= |A| e^{ -i \phi + i \delta}$, where $\phi$ is the CP violating weak phase and $\delta$ is the CP conserving strong phase.

The full phase space for the $\oB\to D^*\ell^-{\bar\nu}$ decay is obtained by varying the kinematic variables over their allowed ranges which are as follows: $m^2_\ell \leq q^2 \leq m^2_B - m^2_{D^*}, 0 \leq \theta_{D^*,\ell} \leq \pi,$ and $0 \leq \chi \leq 2\pi$. One can now construct several observables by integrating the distribution of Eq.~(\ref{eq:angdist}) over one or more of these kinematic variables. The first of these is the differential decay distribution as a function of $q^2$, constructed by integrating over the full range of allowed values for all three helicity angles.
\bea
\frac{d\Gamma}{dq^2} &=& \frac{1}{4}\[3\,I_1^c - I_2^c + 2\,(3\,I_1^s - I_2^s)\].~
\eea
Next, one can construct double-differential decay distributions as functions of $q^2$ and one other angle variable at a time, obtained by integrating over the other two angles.
\bea
\frac{d^2\Gamma}{dq^2d\cos\theta^*} &=& \frac{3}{4}\frac{d\Gamma}{dq^2}\[2\,F_L^{D^*}(q^2)\cos^2\theta^* + F_T^{D^*}(q^2)\sin^2\theta^*\],~ \\
\frac{d^2\Gamma}{dq^2d\cos\theta_\ell} &=& \frac{d\Gamma}{dq^2}\(\frac{1}{2} + A_{FB}\,\cos\theta_\ell + \frac{1 - 3\,{\tilde F}^\ell_L}{4}\,\frac{3\,\cos^2\theta_\ell - 1}{2}\),~ \\
\frac{d^2\Gamma}{dq^2d\chi} &=& \frac{1}{2\pi}\frac{d\Gamma}{dq^2}\(1 + S_3\,\cos2\chi + S_9\,\sin2\chi\) ~, \label{eq:dddd}
\eea
where $F_{L(T)}^{D^*}(q^2)$ is the longitudinal (transverse) polarization of the $D^*$, $A_{FB}$ is the charged-lepton forward-backward asymmetry, and $S_9$ is a triple-product asymmetry. The coefficient functions that appear in Eq.~(\ref{eq:dddd}) can be expressed in terms of the angular coefficients, $I_i^{(s,c)}$, as follows. 
\bea
F_L^{D^*}(q^2) &=& 1 - F_T^{D^*}(q^2) ~=~ \frac{3\,I_1^c - I_2^c}{3\,I_1^c - I_2^c + 2\,(3\,I_1^s - I_2^s)} ,~ \\
A_{FB}(q^2) &=& \frac{3}{2}\,\frac{2\,I_6^s + I_6^c}{3\,I_1^c - I_2^c + 2\,(3\,I_1^s - I_2^s)} ,~ \\
{\tilde F}^\ell_L(q^2) &=& \frac{I_1^c - 3\,I_2^c + 2(I_1^s - 3\,I_2^s)}{3\,I_1^c - I_2^c + 2\,(3\,I_1^s - I_2^s)},~ \\
S_3(q^2) &=& \frac{4\,I_3}{3\,I_1^c - I_2^c + 2\,(3\,I_1^s - I_2^s)} ,~ \\
S_9(q^2) &=& \frac{4\,I_9}{3\,I_1^c - I_2^c + 2\,(3\,I_1^s - I_2^s)} .~
\eea

Note that there are additional observables that can be extracted from data by performing asymmetric integrals over more than one angles. We discuss some such observables in Section \ref{sec:npsig}.

\section{New-Physics Implementation in EvtGen} \label{sec:npimpl}

We implement the preceding discussion in the EvtGen MC simulation framework as the new BTODSTARLNUNP decay model. This NP generator, BTODSTARLNUNP, can run either in a standalone mode or be integrated into a software framework of a $B$-physics experiment. The model includes SM contributions, various NP parameters as well as their interference. The model takes the NP parameters $\delta C_{V_L}\equiv g_L$, $C_{V_R}$, $C_{S_L}$, $C_{S_R}$, and $C_T$ as inputs. The user specifies the NP parameters keeping in mind that the scalar coefficients ($C_{S_L}, C_{S_R}$) are related to each other by Eq.~(\ref{eq:nuconstraint}).  Each of these parameters can take complex values as inputs and are entered in the user decay file. The default value for each parameter has been set to zero so that when no value is specified for these parameters the code returns SM results. Below we present an example of a user decay file to illustrate the usage of the NP MC generator.

\begin{verbatim}
## first argument is cartesian(0) or polar(1) representation of NP coefficients which
## are three consecutive numbers {id, Re(C), Im(C)} or {coeff id, |C|, Arg(C)}
## id==0 \delta C_VL -- left-handed vector coefficient change from SM
## id==1 C_VR -- right-handed vector coefficient
## id==2 C_SL -- left-handed scalar coefficient
## id==3 C_SR -- right-handed scalar coefficient
## id==4 C_T  -- tensor coefficient

Decay B0
## B0 -> D*- e+ nu_e is generated with the Standard Model only
1   D*-    e+   nu_e   BTODSTARLNUNP;
Enddecay

Decay anti-B0
## anti-B0 -> D*+ mu- anti-nu_mu is generated with the addition of New Physics
1   D*+    mu-   anti-nu_mu   BTODSTARLNUNP 0 0 0.06 0 1 0.075 0 2 0 -0.2 3 0 0.2;
Enddecay

End
\end{verbatim}
To generate NP the user inputs several arguments in the user decay file. The first of these specifies whether the remaining arguments are to be entered in Cartesian (0) or polar (1) coordinate system. Next, the user enters sets of three values. The first specifies the type of NP coupling ($\delta C_{V_L}, C_{V_R}, C_{S_L}, C_{S_R},$ and $C_T$), while the second and third represent the real and imaginary parts in Cartesian coordinates, or magnitude and complex phase in polar coordinates.
In the above example we have shown how the user can generate events for the SM as well as for a specific NP scenario which in our case is NP scenario 2. A complete version of the NP MC tool with an implementation of the BTODSTARLNUNP decay model can be found in Ref.~\cite{Campagna:2022evt}.

\section{Signatures of New Physics} \label{sec:npsig}

The ratios of branching fractions as well as the differential $q^2$ distributions have limited sensitivity to NP for $b\to c\ell \nu$, $\ell = e, \mu$, which receive tree-level contributions in the SM and are hence unsuppressed. In contrast, angular observables have much better sensitivity to the interference between SM and NP. The optimal sensitivity to NP can be obtained by studying these angular observables as functions of $q^2$. We will examine four angular asymmetries as functions of $q^2$ to make predictions for our NP scenarios, $A_{FB}$, $S_3$, $S_5$, and $S_7$. $A_{FB}$ and $S_3$ are previously defined in Section \ref{sec:th}, while $S_5$ and $S_7$ are the coefficients of $\sin\theta_\ell\sin2\theta^*\cos\chi$ and $\sin\theta_\ell\sin2\theta^*\sin\chi$, respectively. These asymmetries can be constructed from the full angular distribution of Eq.~(\ref{eq:angdist}) through asymmetric integrals shown below.
\bea
A_{FB}(q^2) &=& \(\frac{d\Gamma}{dq^2}\)^{-1}\[\int\limits_0^1 - \int\limits_{-1}^0\]d\cos\theta_\ell\,\frac{d^2\Gamma}{d\cos\theta_\ell dq^2} ,~ \label{eq:AFB} \\
S_3 (q^2) &=& \(\frac{d\Gamma}{dq^2}\)^{-1}\[\int\limits_0^{\pi/4} - \int\limits_{\pi/4}^{\pi/2} - \int\limits_{\pi/2}^{3\pi/4} + \int\limits_{3\pi/4}^{\pi} + \int\limits_{\pi}^{5\pi/4} - \int\limits_{5\pi/4}^{3\pi/2} - \int\limits_{3\pi/2}^{7\pi/4} + \int\limits_{7\pi/4}^{2\pi}\]d\chi\,\frac{d^2\Gamma}{dq^2 d\chi} ,~ \label{eq:S3} \\
S_5(q^2) &=& \(\frac{d\Gamma}{dq^2}\)^{-1}\[\int\limits_0^{\pi/2}-\int\limits_{\pi/2}^\pi - \int\limits_\pi^{3\pi/2} + \int\limits_{3\pi/2}^{2\pi}\]d\chi\[\int\limits_0^1 - \int\limits_{-1}^0\]d\cos\theta^*\,\frac{d^3\Gamma}{dq^2d\cos\theta^*d\chi} ,~\label{eq:S5} \\
S_7(q^2) &=& \(\frac{d\Gamma}{dq^2}\)^{-1} \[\int\limits_0^\pi - \int\limits_\pi^{2\pi}\] d\chi \[\int\limits_0^1 - \int\limits_{-1}^0\]d\cos\theta^*\,\frac{d^3\Gamma}{dq^2d\cos\theta^*d\chi} ~\label{eq:S7}.~
\eea

To extract these asymmetries from data, we calculate the integrals in Eqs.~(\ref{eq:AFB}-\ref{eq:S7}) from binned distributions of the appropriate angular variables. For example, consider $S_5$. This distribution involves asymmetric integrals over both $\cos\theta^*$ and $\chi$. For a given bin of $q^2$, we first divide the events into $\chi$ bins of size $\pi/2$. In each of these bins, we then divide the events into $\cos\theta^*$ bins of size 1. This gives us 8 bins corresponding to the various terms of Eq. (\ref{eq:S5}), which we will label $N_i$ with $i=1,2,...,8$. To find the value of $S_5$ for a given $q^2$ bin, we then combine the $N_i$'s in the same way as the integrals in Eq. (\ref{eq:S5}), normalized by $\sum\limits_{i=1}^8N_i$.
\begin{table}[t!]
\renewcommand{\arraystretch}{1.2}
\begin{tabular}{|c|c|c|c|} \hline
Observable & Angular Function & NP Dependence & $m_\ell$ suppression order \\ \hline
\multirow{7}{*}{$A_{FB}$} & \multirow{7}{*}{$\cos\theta_\ell$} & $\Re\[g_Tg_P^*\]$ & \multirow{2}{*}{$\cO(1)$} \\ 
&& $\Re\[(1+g_L-g_R)(1+g_L+g_R)^*\]$ & \\ \cline{3-4} 
&& $\Re\[(1+g_L-g_R)g_P^*\]$ & \multirow{3}{*}{$\cO(m_\ell/\sqrt{q^2})$} \\ 
&& $\Re\[g_T(1+g_L-g_R)^*\]$ & \\ 
&& $\Re\[g_T(1+g_L+g_R)^*\]$ & \\ \cline{3-4} 
&& $|1+g_L-g_R|^2$ & \multirow{2}{*}{$\cO(m^2_\ell/q^2)$} \\
&& $|g_T|^2$ & \\\hline
\multirow{3}{*}{$S_3$} & \multirow{3}{*}{$\sin^2\theta^*\sin^2\theta_\ell\cos2\chi$} & $|1 + g_L +g_R|^2$ & \multirow{3}{*}{$\cO(1),~\cO(m^2_\ell/q^2)$} \\ 
&& $|1 + g_L - g_R|^2$ & \\
&& $|g_T|^2$ & \\ \hline
\multirow{6}{*}{$S_5$} & \multirow{6}{*}{$\sin2\theta^*\sin\theta_\ell\cos\chi$} & $\Re\[g_Tg_P^*\] $ & $\cO(1)$ \\ \cline{3-4}
&& $|1 + g_L - g_R|^2$ &  $\cO(1),~\cO(m^2_\ell/q^2)$ \\\cline{3-4} 
&& $\Re\[(1+g_L-g_R)g_P^*\]$ &\\
&& $\Re\[g_T(1+g_L-g_R)^*\]$ & $~\cO(m_\ell/\sqrt{q^2})$\\
&& $\Re\[g_T(1+g_L+g_R)^*\]$ &\\\cline{3-4} 
&& $|g_T|^2$ & $~\cO(m^2_\ell/q^2)$\\ \hline
\multirow{4}{*}{$S_7$} & \multirow{4}{*}{$\sin2\theta^*\sin\theta_\ell\sin\chi$} & $\Im\[g_Pg_T^*\] $ & $\cO(1)$ \\\cline{3-4} 
&& $\Im\[(1+g_L+g_R)g_P^*\]$ & \multirow{2}{*}{$\cO(m_\ell/\sqrt{q^2})$}\\
&& $\Im\[(1+g_L-g_R)g_T^*\]$ & \\\cline{3-4} 
&& $\Im\[(1+g_L-g_R)(1+g_L+g_R)^*\]$ & $\cO(m^2_\ell/q^2)$\\ \hline
\end{tabular}
\caption{Angular functions corresponding to angular observables $A_{FB}, S_3, S_5$, and $S_7$ alongside NP parameters that contribute to each. The dependence on NP parameters has been separated into different orders of $m_\ell/\sqrt{q^2}$.}
\label{tab:S-dependencies}
\renewcommand{\arraystretch}{1}
\end{table}

When generating our predictions, we used $\Delta A_{FB} = A_{FB}(B\to D^{*} \mu\nu) - A_{FB} (B\to D^{*} e \nu)$, $\Delta S_3 = S_3(B\to D^{*} \mu\nu) - S_3 (B\to D^{*} e \nu)$, and $\Delta S_5 = S_5(B\to D^{*} \mu\nu) - S_5 (B\to D^{*} e \nu)$, where the electron mode has been generated with the SM only while the muon mode contains both SM and NP contributions. These are $\Delta$-type observables as defined above, which eliminate most of the QCD uncertainties in the form factors, allowing for a clean measurement of LFUV NP. The asymmetry $S_7$ is always zero in the SM, and therefore was not recast into the form of a $\Delta$ observable. The NP dependences of $A_{FB}$, $S_3$, $S_5$, and $S_7$ are given in Table \ref{tab:S-dependencies}. Note that these dependencies have different weights, which are dependent on $q^2$. For all theory plots presented here, we have only used uncorrelated central values of the form factor parameters as listed in Tables.~\ref{tab:inputs-others} and \ref{tab:FFparam}. We verify that the $\Delta$ variables have minimal dependence on form factors. As a test, we consider BGL \cite{Boyd:1997kz}, CLN \cite{Caprini:1997mu},  and HQET \cite{Bordone:2019vic} form factor parameterizations. There are also other form factor models \cite{Faustov:2022ybm,Bernlochner:2022ywh}. Unless otherwise stated, we use the CLN parameterization of the hadronic form factors as the default in our plots.

\section{New-Physics Sensitivity and Results} \label{sec:npsens}

\begin{figure}[t]
    \centering
    \includegraphics[scale=0.47]{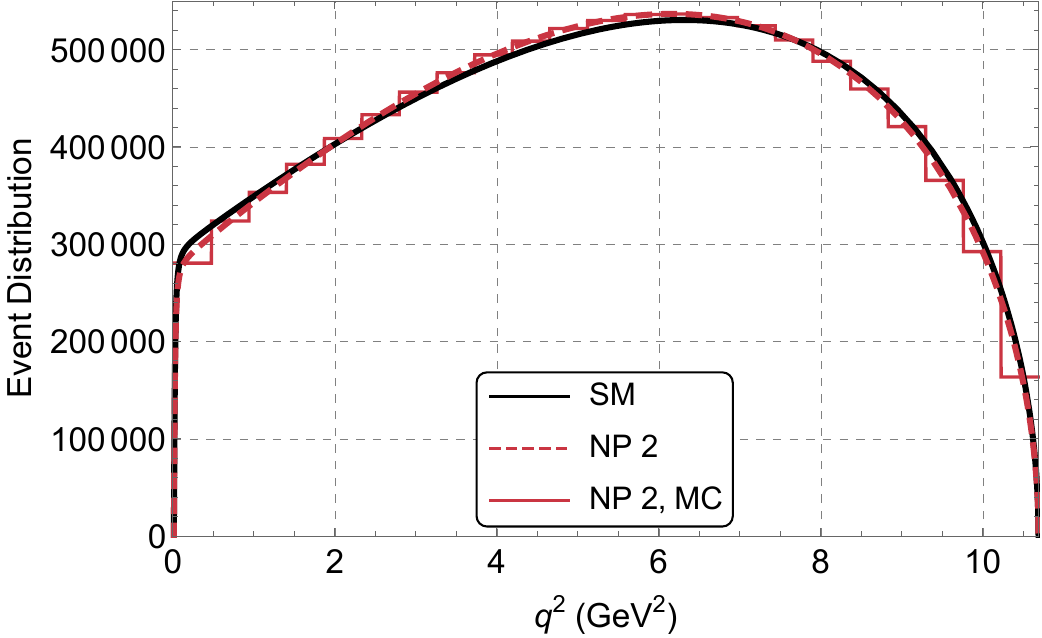}~~
    \includegraphics[scale=0.47]{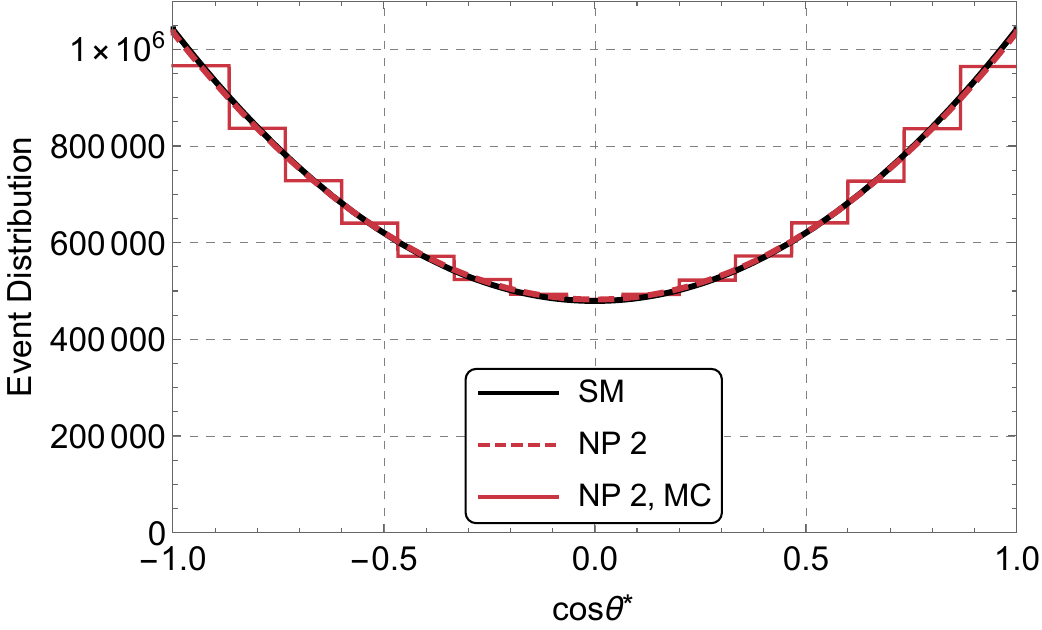}~~\\
    \includegraphics[scale=0.47]{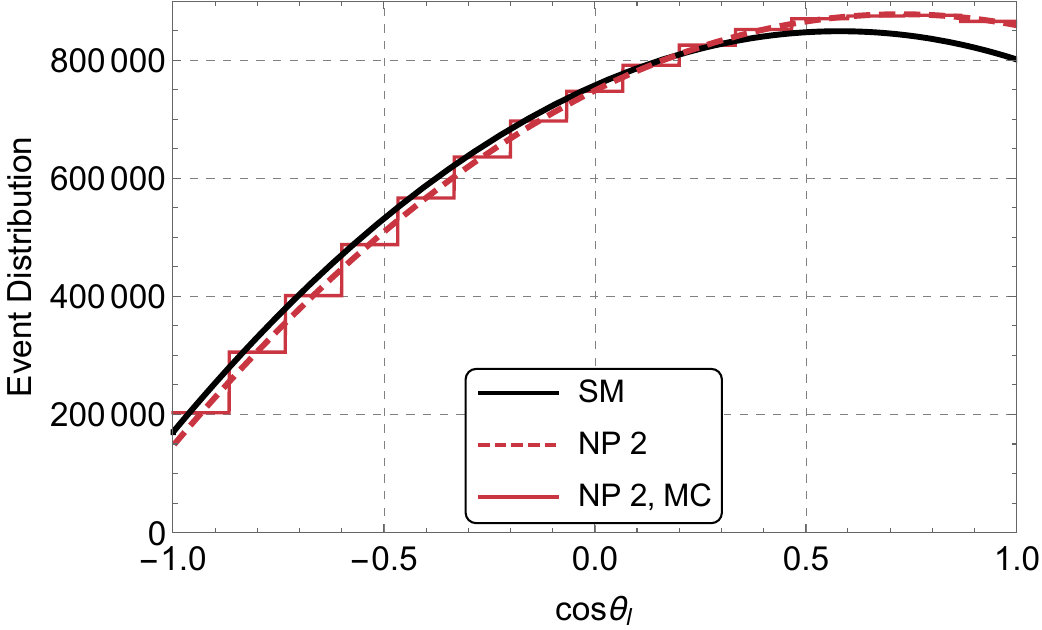}~~
    \includegraphics[scale=0.47]{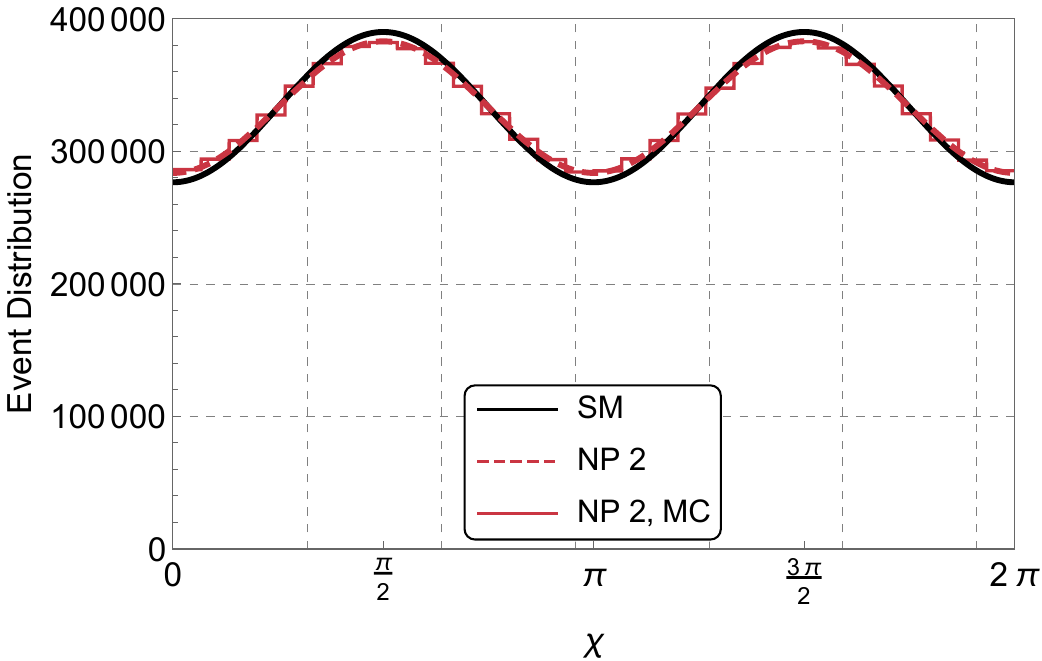}~~
    \caption{Distribution of $\oB\to D^*\ell^-{\bar\nu}$ events as functions of (clockwise from top left) $q^2$, $\cos\theta^*$, $\chi$, and $\cos\theta_\ell$. Theory predictions are shown for the SM (solid black curve) and for NP Scenario 2 (dashed red curve). EvtGen data are shown for NP Scenario 2 (solid red histogram). Each plot is fully integrated over three of the four kinematic variables. The $q^2$ range is divided into 23 equal bins, to reflect the expected resolution of experimental measurements. The angular bins are chosen to be sufficiently fine to compare MC data to the theory. The $\cos\theta$ ranges are divided into 15 equal bins, and the $\chi$ range, being twice as large as the $\theta$ ranges, is divided into twice as many bins.}
    \label{fig:distribs}
\end{figure}

The $q^2$ distribution alone has little sensitivity to NP, as shown in Fig.~\ref{fig:distribs}. On the other hand, angular asymmetries as functions of $q^2$ are quite sensitive to NP couplings. In particular, the angular asymmetries in the angle $\theta_\ell$ and $ \chi$ can be promising probes of NP as shown in Fig.~\ref{fig:distribs}. In this figure, we have used the CLN parameterization to test that our Monte Carlo generator correctly implements the theoretical expressions. However, the angular asymmetries remain quite sensitive to form-factor uncertainties. As an example, the uncertainty in the predictions for $A^\mu_{FB}$ in the SM with four different form-factor parameterizations is shown below.
To address this issue we consider differences between angular asymmetries in the muon and electron channels using $\Delta$ observables. Later in this section, using $\Delta A_{FB}$ as an example, we show that
the predictions for the $\Delta$ observables are robust against form factor uncertainties using the same four form factor parameterizations. In the SM the form factor uncertainties cancel effectively in the $\Delta $ observables while with NP the cancellation is slightly less effective as the NP violates lepton universality. 

\begin{figure}[htp!]
    \centering
    \includegraphics[scale=0.5]{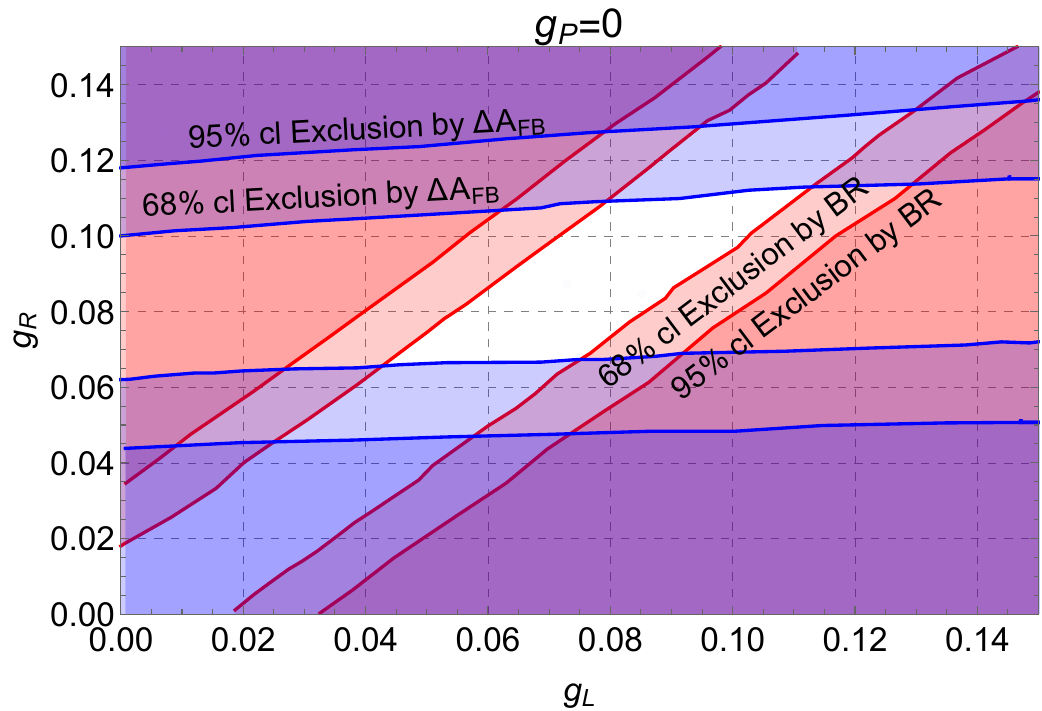}
    \includegraphics[scale=0.5]{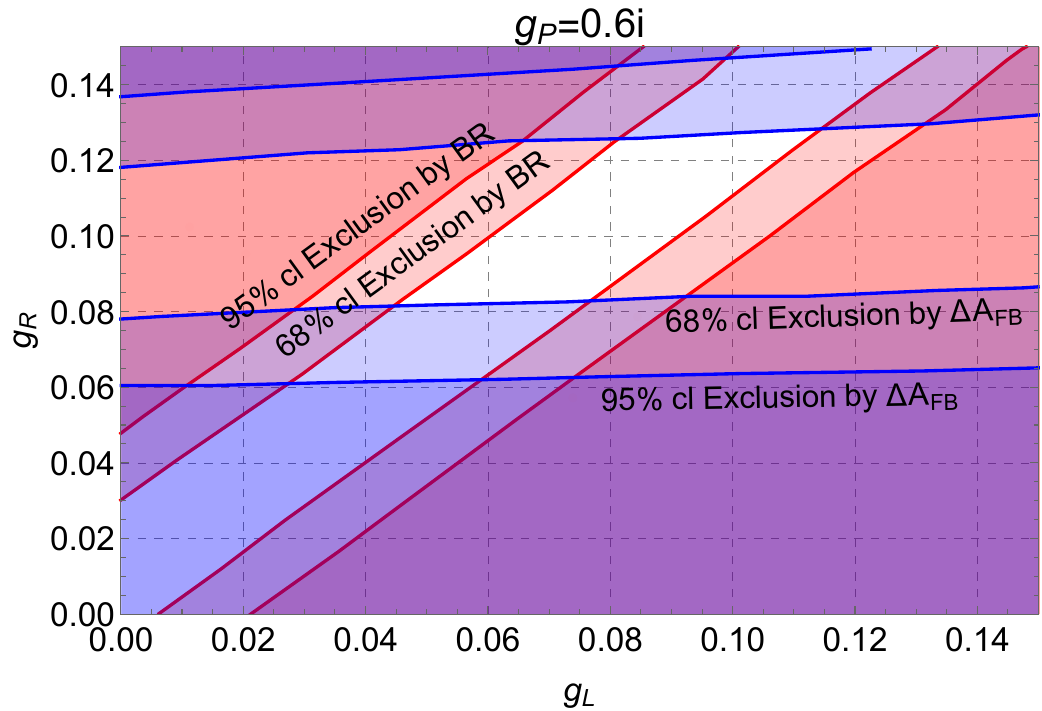}
    \caption{Allowed parameter space in $g_L$ and $g_R$, with $g_P=0$ and $0.6i$. The two constraints used are that the branching ratio of the muon and electron modes must be unity within 3\%, and $\Delta A_{FB}$ must be consistent with the value found in Ref.~\cite{Bobeth:2021lya}. Non-zero values of $g_P$ produce similar plots, with the allowed region in $g_R$ shifting upwards. This exercise also showed that imaginary values of $g_L$ and $g_R$ are not consistent with these constraints.}
    \label{fig:param_plot}
\end{figure}

From our initial scan, we cannot reproduce the experimental-$\Delta A_{FB}$ anomaly with a single NP coupling. Instead, we consider scenarios with several NP couplings.  In order to match 
$\Delta A_{FB}$ from Ref.~\cite{Bobeth:2021lya}, we require a $g_R$ NP coupling. In order to maintain the LFU BR constraint we also need to add a $g_L$ NP coupling that is comparable to $g_R$. In addition, it is also possible to include a $g_P$ contribution, but in order to satisfy the constraints it must be imaginary. We also found that negative or complex values for $g_L$ and $g_R$ are ruled out by these constraints. Fig.~\ref{fig:param_plot} shows the region of parameter space in the $g_L$-$g_R$ plane that is excluded by $\frac{\mathcal{B}(B \to D^* \mu \nu)}{\mathcal{B}(B \to D^* e \nu)} = 1.00 \pm 0.03~(0.06)$ in red and the region in blue excludes $\Delta A_{FB} =0.0349 \pm 0.0089~(0.0178)$ when the error is taken in the $68\%~(95\%)$ C.L. Further, we observe that an additional non-zero imaginary pseudoscalar interaction strength produces an upward shift in the allowed region of $g_R$ while $g_L$ remains almost the same as shown in the right plot of Fig.~\ref{fig:param_plot}. In this section we provide results corresponding to the three distinct NP Scenarios indicated in Table \ref{tab:scenarios} chosen with the above constraints. 

\begin{table}[htp!]
\begin{tabular}{l c c c} \hline
& $g_L$ & $g_R$ & ~$g_P$ \\ \hline
Scenario 1:~~ & ~$0.06$~~ & ~$0.075$~ & ~~$0.2i$~ \\
Scenario 2:~~ & ~$0.08$~~ & ~$0.090$~ & ~~$0.6i$~ \\
Scenario 3:~~ & ~$0.07$~~ & ~$0.075$~ & ~~$0$~     \\ \hline
\end{tabular}
\caption{Values of NP coefficients for three distinct NP scenarios considered in this paper and used for generating the results presented in this section. \label{tab:scenarios}}
\end{table}

\begin{figure}[htbp!!]
    \centering
    \includegraphics[scale=0.5]{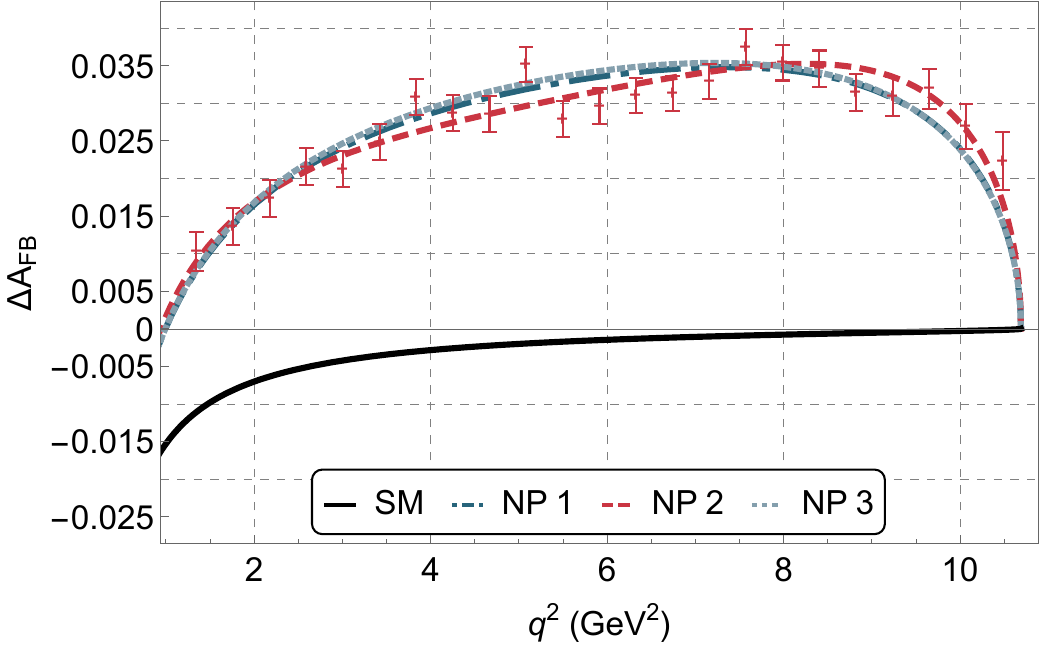}~~
    \includegraphics[scale=0.5]{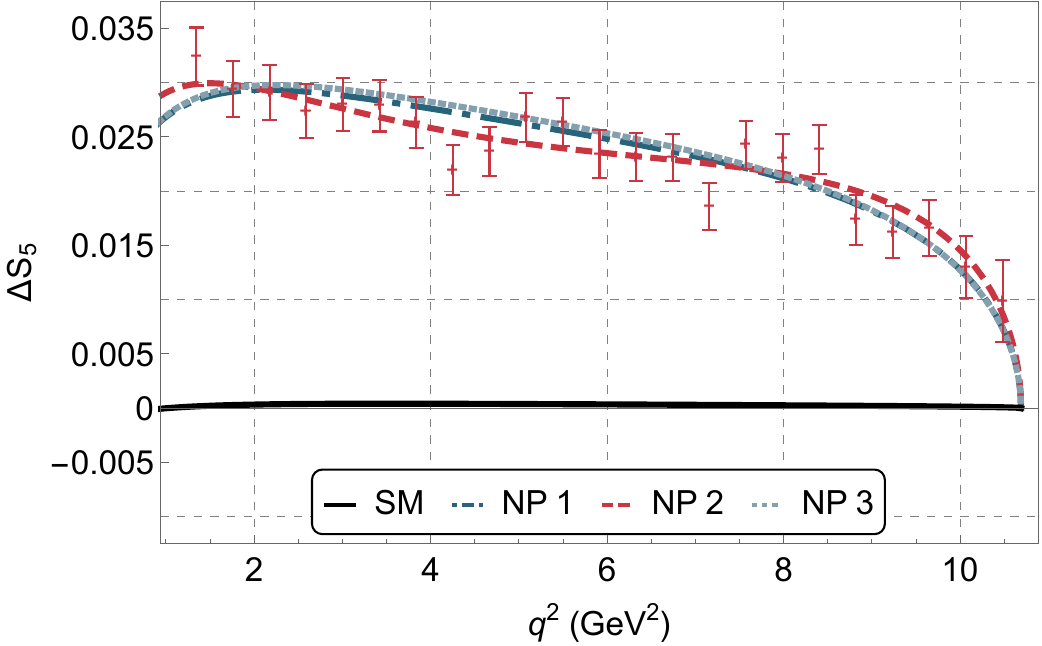}~~\\
    \includegraphics[scale=0.5]{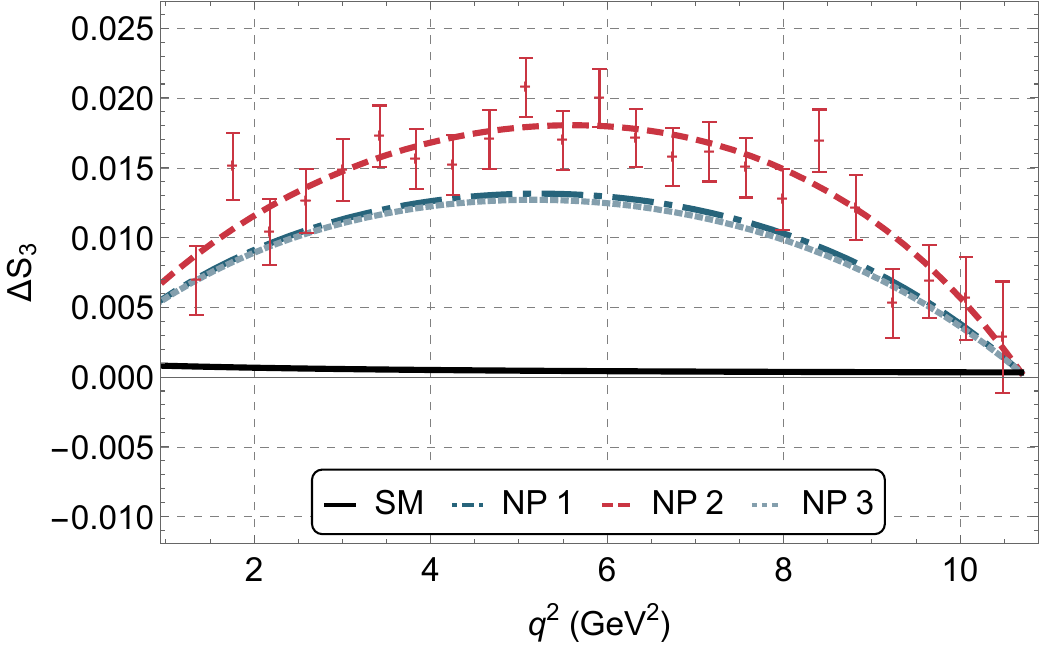}~~
    \includegraphics[scale=0.5]{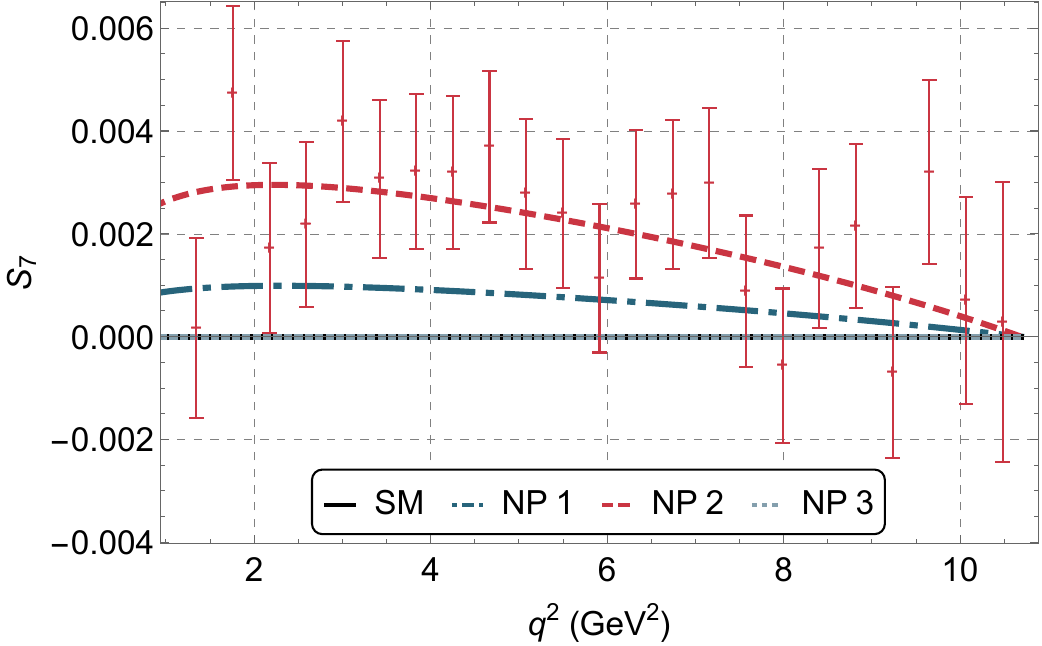}
    \caption{$\dAFB$, $\Delta S_5$, $\Delta S_3$, and $S_7$ plotted as functions of $q^2$ for different values of NP coefficients. Here we have used the CLN parameterizations of the FFs. The NP parameters were chosen so that the ratio of semi-leptonic branching fractions is constrained to be within $3\%$ of unity, as well as the $\Delta A_{FB}$ for the full $q^2$ range is within the interval $0.0349 \pm 0.0089$. EvtGen data for NP Scenario 2 ($g_L=0.08$, $g_R=0.09$, $g_P=0.6i$) generated with $10^7$ events (anticipated Belle II statistics) are shown as points with error bars. Theory curves are presented for all three NP Scenarios: Scenario 1 is dot-dashed blue, Scenario 2 is dashed red and Scenario 3 is dotted blue.}
    \label{fig:observables}
\end{figure}

\begin{figure}[htp!]
    \centering
    \includegraphics[scale=0.5]{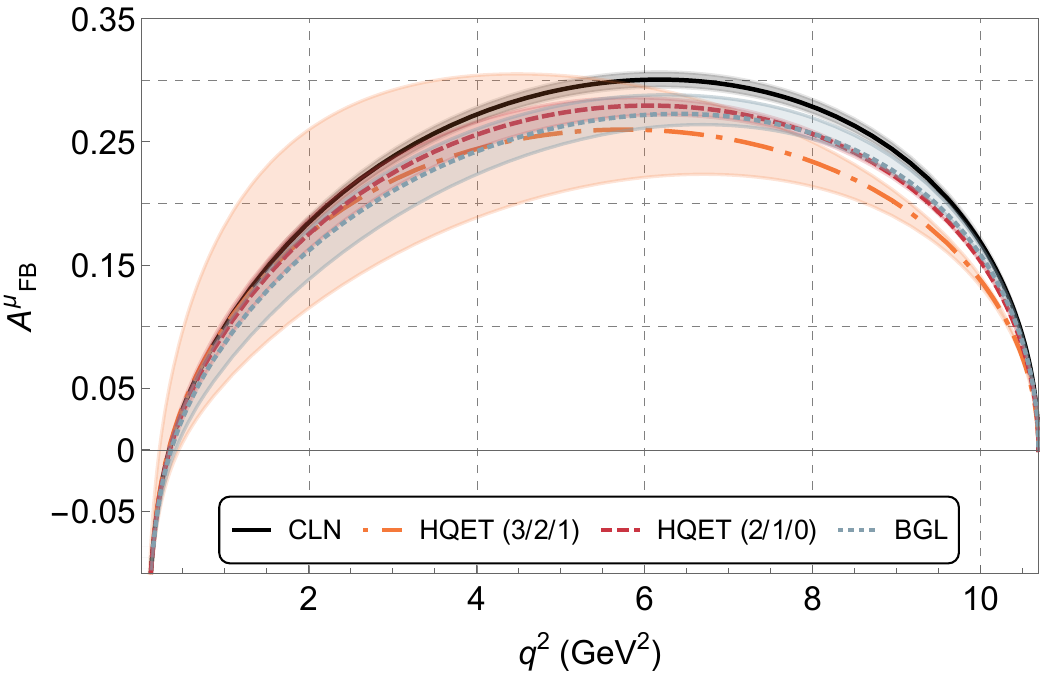}
    \includegraphics[scale=0.5]{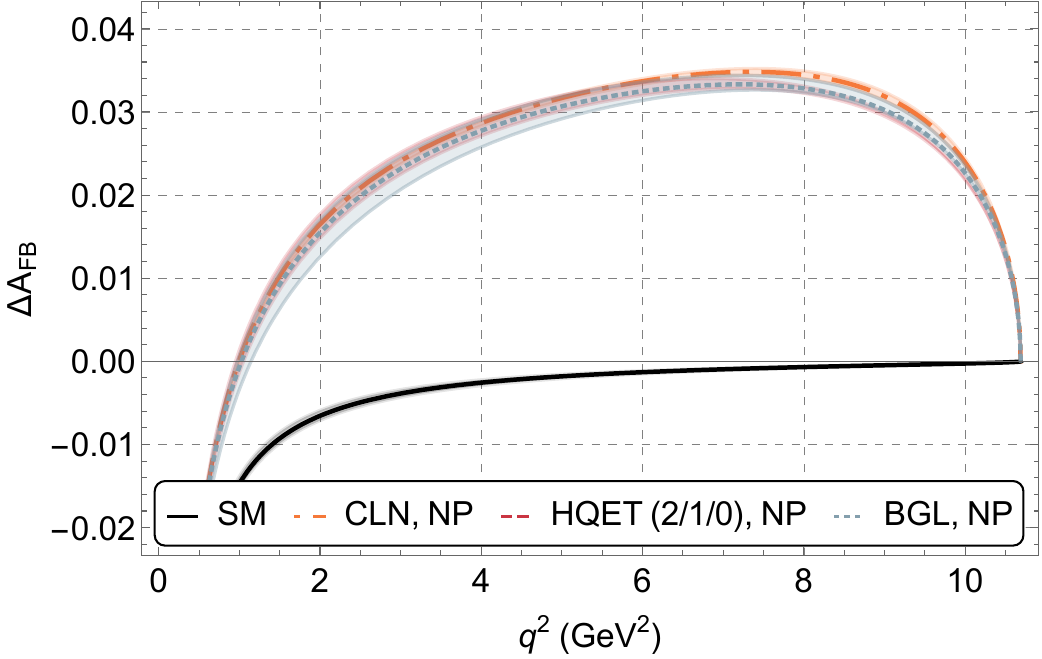}
    \caption{${A^\mu_{FB}}$ (left plot) in the SM and $\Delta A_{FB} = A^\mu_{FB} - A^e_{FB}$ (right plot) for different form factor parameterizations. The left plot shows the SM predictions for various form factor parameterizations, while the right plot demonstrates the effects of form factor uncertainties on $\Delta A_{FB}$ in NP Scenario 1 ($g_L=0.06$, $g_R=0.075$, and $g_P=0.2i$). The solid black curve in the right plot represents the SM prediction for both CLN and HQET (2/1/0) parameterizations. Note that the vertical scale of the right plot is approximately a factor of ten smaller than that of the left plot. Note also the large negative value at the low $q^2$ limit. A cutoff of 1.14 GeV$^2$ is chosen to avoid this. Note that for SM and NP 3, $\langle S_7\rangle$ is exactly zero and are not distinguishable.}
    \label{fig:CLN_HQET_BGL}
\end{figure}

To optimize sensitivity, it is important to measure the $\Delta$ observables as functions of $q^2$. Using the benchmark scenarios above, we show in Fig.~\ref{fig:observables} the predictions for the $\Delta$ observables. As discussed earlier these observables are sensitive to NP couplings and have much reduced dependence on form-factor uncertainties. In the figure, the SM expectations for these quantities are shown using solid black curves. In addition to the two $\Delta$ observables, $\Delta A_{FB}$ and $\Delta S_5$, Fig.~\ref{fig:observables} also shows the $q^2$ dependence of the observable $\Delta S_3$ and $S_7$. $S_7$ represents an angular asymmetry in $\sin\chi$, where $\chi$ is the azimuthal angle between the decay planes. This is a CP-odd triple-product asymmetry, which is predicted to be identically zero in the SM for any $q^2$. We find that NP scenarios with an imaginary $g_P$ are able to produce a small non-zero signal in the $q^2$ distribution of $S_7$ as shown in Fig.~\ref{fig:observables}. 

The observable $S_3$ is the coefficient of $\cos 2\chi$ term in the angular distribution and can be extracted using the asymmetric integral defined in Eq.~\eqref{eq:S3}. Although $\Delta S_3$ is close to zero in the SM, NP can produce a non-zero $\Delta S_3$ in the $q^2$ range  as shown in the lower left plot of Fig.~\ref{fig:observables}. In Fig.~\ref{fig:CLN_HQET_BGL}, using $\Delta A_{FB}$ as an example, we show that the predictions for the $\Delta$-observables are largely independent of form factor parametrization and the uncertainties of the form factor parameters.

Note that due to lepton mass and helicity effects, $\Delta A_{FB}$ is negative in the low $q^2$ region even in the SM. In fact, at the lower momentum transfer threshold, i.e in the limit $q^2 \to m_\ell^2$, the forward-backward asymmetry $A_{FB}^\ell \to -1$ which is seen as a large dip in the $q^2$ distribution as shown in Fig.~\ref{fig:CLN_HQET_BGL}. Hence, for the best experimental sensitivity to NP, we advocate a necessary low $q^2$ cut of 1.14 GeV$^2$ on such observables in order to predict them unambiguosly. 
%We find that the SM predictions for $\Delta A_{FB}$ over the full $q^2$ range is $-5.4 \times 10^{-3}$, and with a low-$q^2$ cutoff of 1.14 GeV$^2$ it is $-2.5 \times 10^{-3}$. 

\begin{table}[bp!]
\begin{tabular}{l c c c c} \hline
 & $\langle \Delta A_{FB}\rangle $ & $\langle \Delta S_3\rangle$ & $\langle \Delta S_5 \rangle$ & $\langle S_7 \rangle$ \\ &\% & \% & \% & $\times 10^{-3}$\\\hline
SM: ~~ & -0.23$\pm0.02$ & 0.052$^{+0.004}_{-0.002}$ & 0.044$\pm0.005$ & 0 \\
NP 1:~~ & ~2.7$\pm0.1$~~ & ~0.87$^{+0.12}_{-0.07}$~ & ~2.21$^{+0.08}_{-0.09}$~ & 0.56$^{+0.03}_{-0.04}$\\
NP 2:~~ & ~2.8$\pm0.1$~~ & ~1.27$^{+0.13}_{-0.09}$~ & ~2.25$^{+0.08}_{-0.10}$~ & 1.69$^{+0.09}_{-0.10}$\\
NP 3:~~ & ~2.8$\pm0.1$~~ & ~0.83$^{+0.12}_{-0.04}$~  & ~2.24$^{+0.08}_{-0.09}$~ & 0\\ \hline
\end{tabular}
\caption{Theoretical predictions of integrated $\Delta A_{FB}$, $\Delta S_3$, $\Delta S_5$, and $S_7$ for SM and each NP scenario using the BGL form factor parameterization with estimated theoretical uncertainties. Note that for SM and NP 3, $\langle S_7\rangle$ is exactly zero as all associated couplings are real.}
\label{tab:predictions}
\end{table}

\begin{figure}[htp!]
    \centering
    \includegraphics[scale=0.55]{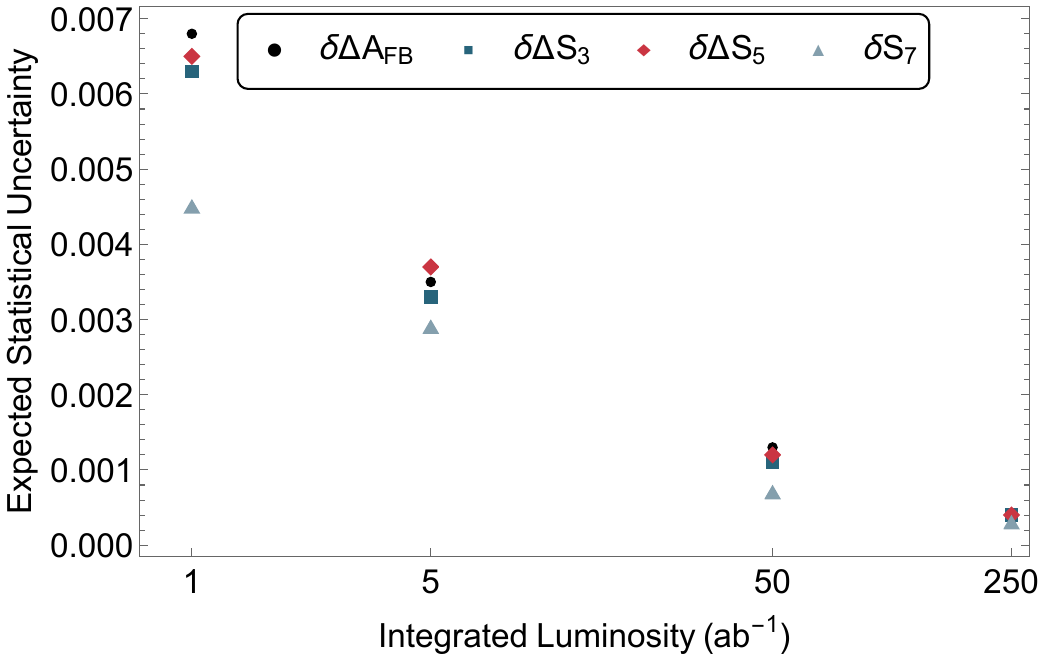}
    \caption{Expected statistical uncertainties for the four observables at 1, 5, 50, and 250 ab$^{-1}$ of Belle II data. These expected uncertainties were found using the BTODSTARLNUNP MC simulation.}
    \label{fig:stat_uncertainty}
\end{figure}

\begin{figure}
    \centering
    \includegraphics[scale=0.45]{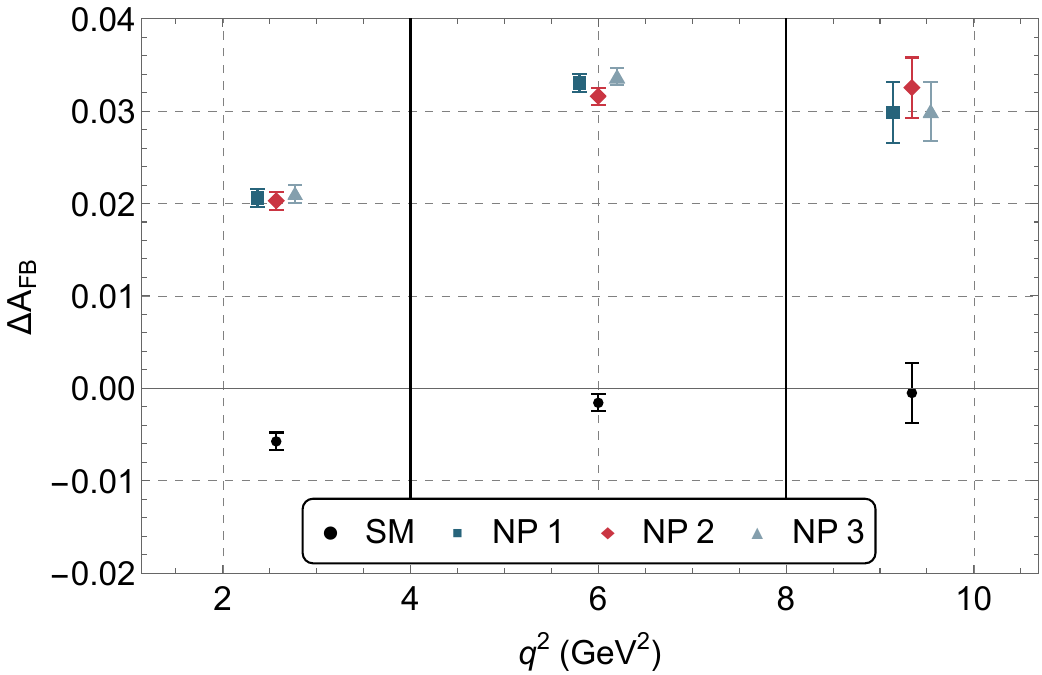}
    \includegraphics[scale=0.45]{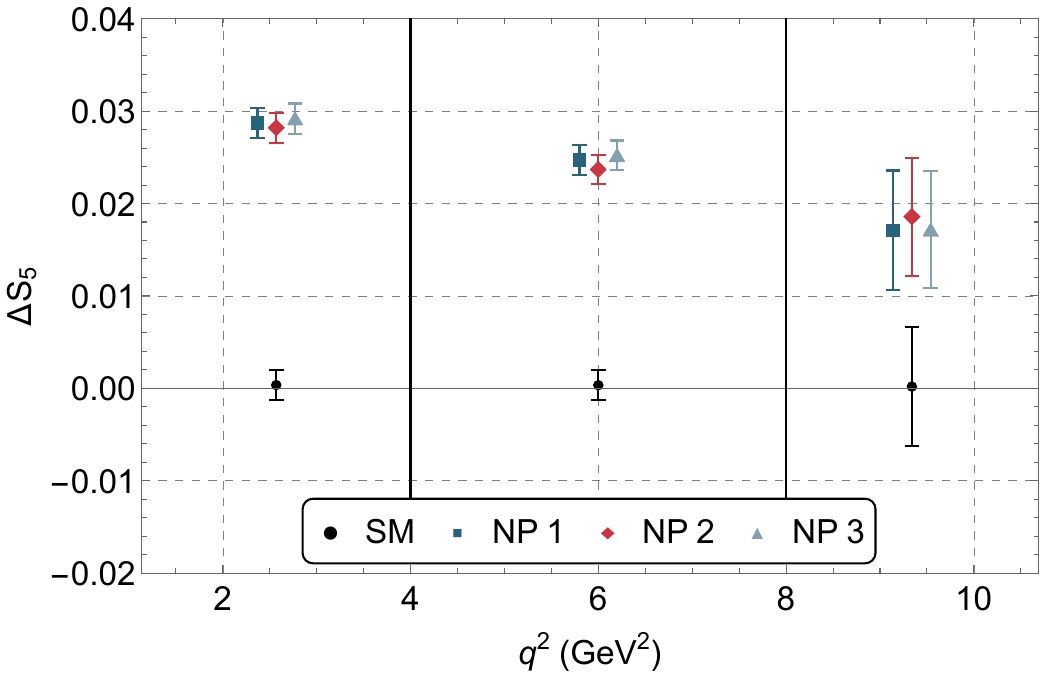}
    \caption{Coarse-binned distributions of $\Delta A_{FB}$ and $\Delta S_5$ versus $q^2$. The horizontal axis spans the allowed range for $q^2$ which has been divided into three bins. The vertical lines at 4 and 8 GeV$^2$ indicate the other edges of these bins. The central values are calculated from theory, and the error bars indicate statistical uncertainties taken from MC simulation with an integrated luminosity of 50 ab$^{-1}$. The NP1 and NP3 predictions have been offset from the center of each bin for clarity.}
    \label{fig:coarsebin}
\end{figure}

\begin{figure}[htp!]
    \centering
    \includegraphics[scale=0.4]{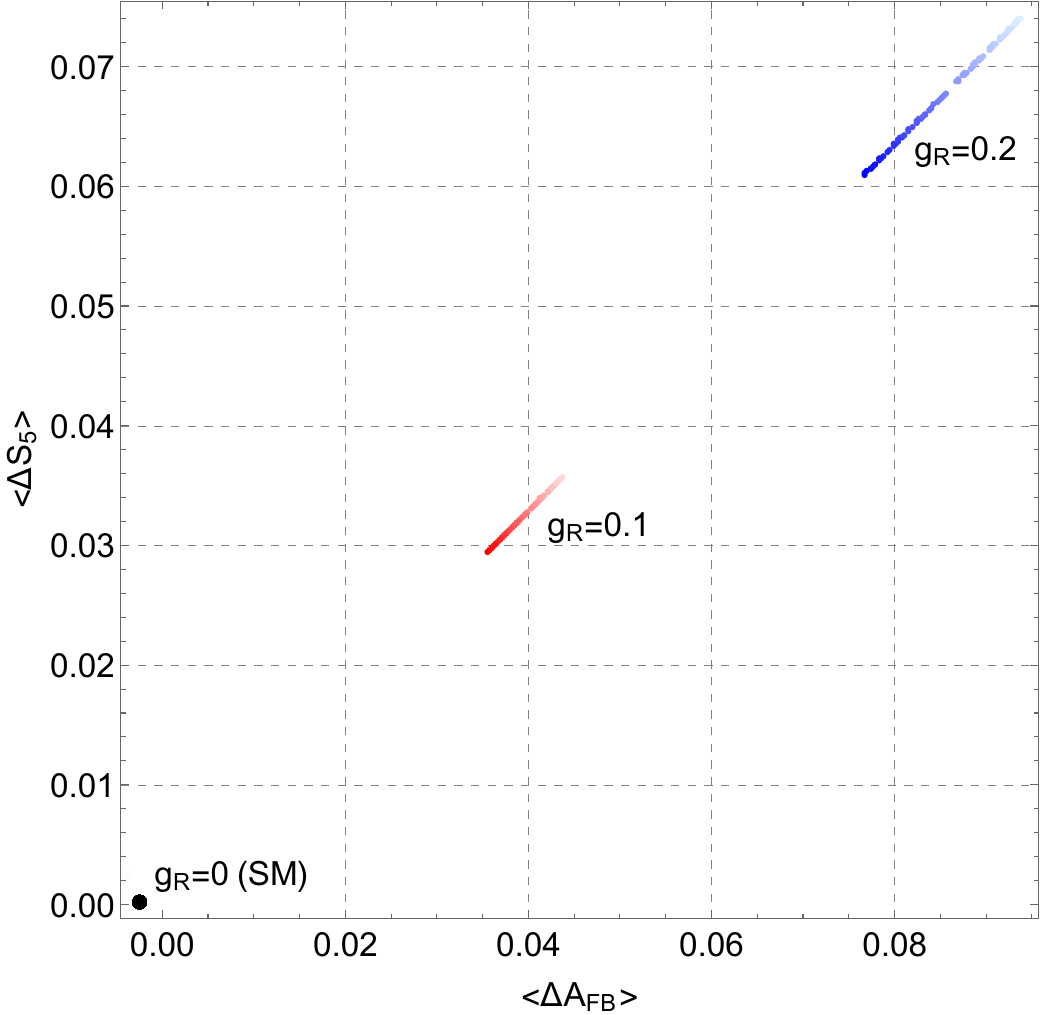}
    \includegraphics[scale=0.41]{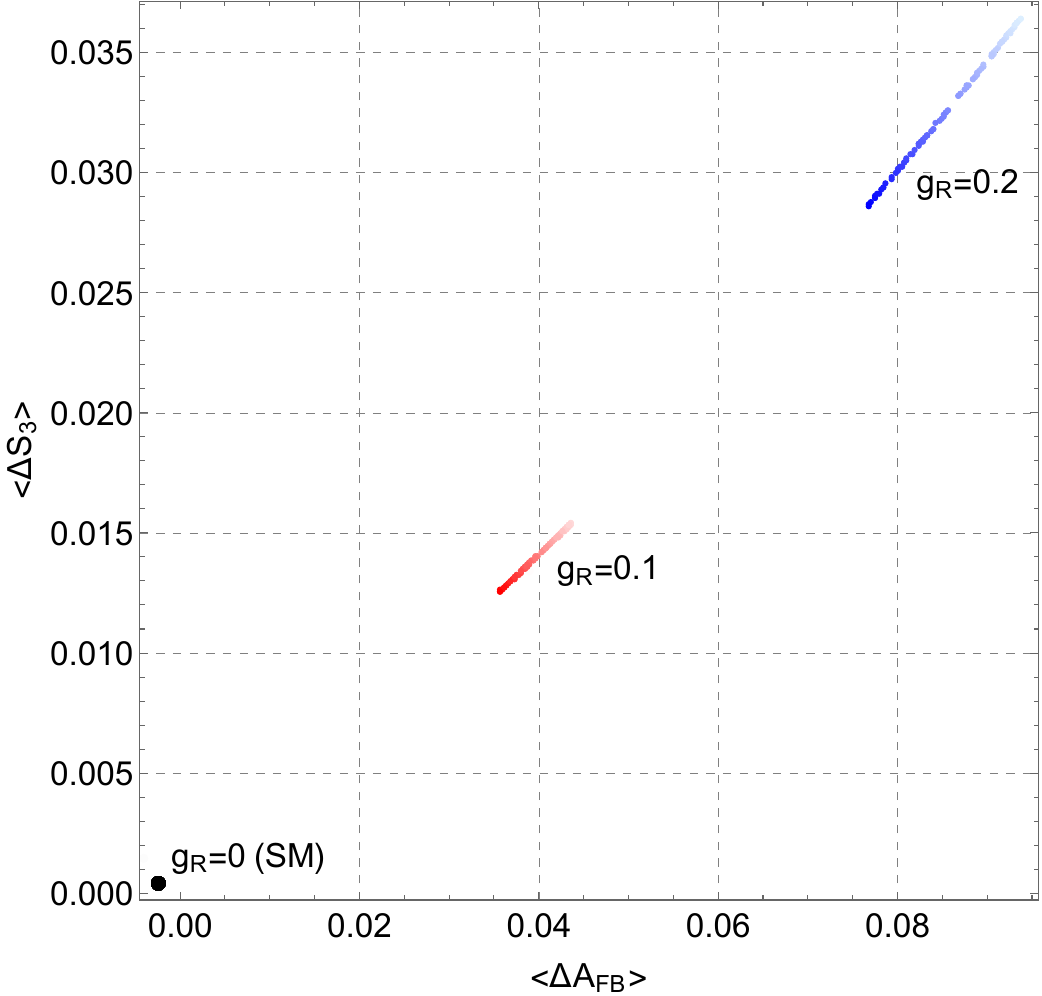}\\
    \includegraphics[scale=0.6]{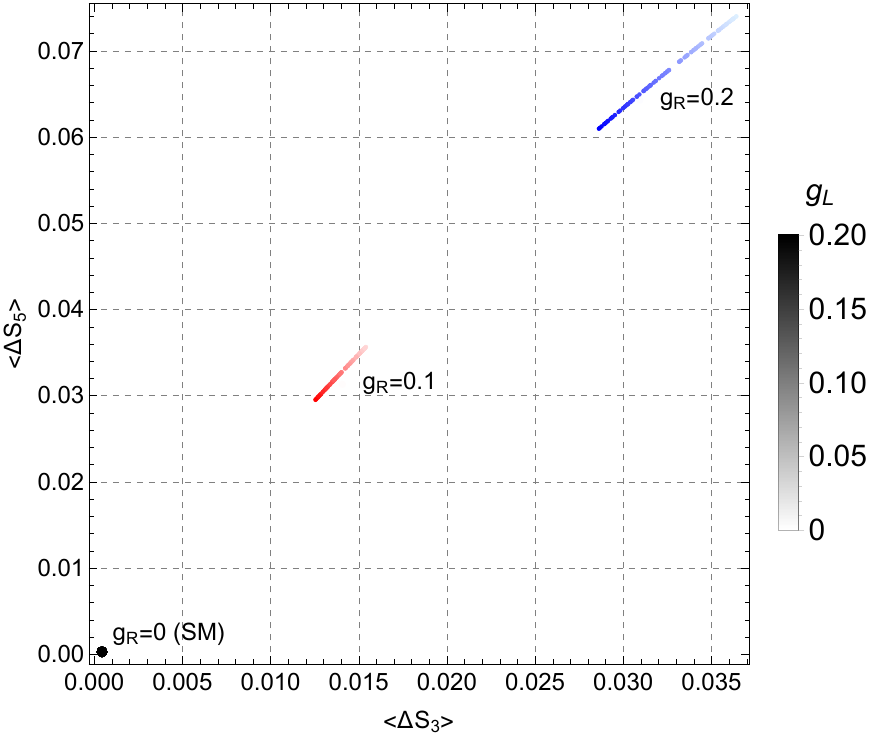}
    \caption{Correlations between $\langle \Delta A_{FB}\rangle$, $\langle\Delta S_3\rangle$, and $\langle\Delta S_5\rangle$ in NP scenarios. For each point, $g_L$ is varied between 0 and 0.2 (light to dark in the color scale as depicted in the bar legend; applies for each value of $g_R$), with $g_R$ = 0, 0.1, or 0.2, which are representative values in the allowed range, and $g_P$ = 0. All points for which only $g_L$ is non-zero return the SM values of the three observables.}
    \label{fig:correlation}
\end{figure}

In addition, in order to improve systematic uncertainties from lepton identification efficiencies, we recommend using the same laboratory momentum cutoff for both $\ell = e$ and $\mu$ channels (see for example \cite{Herren:2022spb}). In order to define the detector acceptance we will represent the magnitude of the transverse momentum of particle $x$ in the lab frame by $|\vec{p}_{T,x}|$ and the ratio of the $z$-component of the momentum over the total momentum as $\cos \alpha$. We use the Belle II acceptances of $|\vec{p}_{T,\ell}| > 0.8$ GeV for the lepton momenta, $|\vec{p}_{T,\pi}| > 0.1$ GeV for the slow pion momenta, and $-0.866 < \cos\alpha < 0.956$ for all final state particles. The theoretical predictions and uncertainties for these observables integrated over the range of $q^2 \in \[1.14~\text{GeV}^2,(m_B-m_{D^*})^2\]$  using the BGL parameterization are displayed in Table \ref{tab:predictions} both for the SM and the specific NP scenarios listed in Table \ref{tab:scenarios}. One can see that the theoretical uncertainties are less than $\sim 5\%$ for both the SM and NP predictions of all integrated observables except $\langle\Delta S_3 \rangle$ which has a~$\sim 15\%$ uncertainty. We also show the variation of the expected statistical uncertainties as a function of the total integrated luminosity for present and future experimental datasets in Fig.~\ref{fig:stat_uncertainty} using MC simulations.

Initially, experiments will measure integrated $\Delta$ observables. As statistics improve, they will proceed to coarse-binned measurements, as shown for example, in Fig.~\ref{fig:coarsebin}. At high statistics, unbinned fits to angular observables will be performed, as shown for example, in Fig.~\ref{fig:observables}.

Furthermore, from Fig.~\ref{fig:observables} we see that NP couplings produce correlated signatures of deviations from the SM in multiple $\Delta$ observables, such as $\Delta A_{FB}$ and $\Delta S_5$. As shown in Fig.~\ref{fig:correlation}, the size of the effect on $\Delta$-observables is determined primarily by $g_R$. In this plot, we have varied the NP parameter $g_L$ between $0$ and $0.2$ for fixed values of $g_R$. In the presence of NP there are strong correlations between the $\Delta$-observables $\Delta A_{FB}$, $\Delta S_5$, and $\Delta S_3$. Therefore, if an experimental signal in $\Delta A_{FB}$ is observed, it should be accompanied by an observation of non-zero $\Delta S_5$ and $\Delta S_3$. Conversely, if a non-zero $\Delta S_5$ is observed, there must also be a non-zero $\Delta A_{FB}$. In the absence of a tensor coupling, a correlation with $\Delta S_3$ is also required.

For the benchmark scenarios described above, we have also checked the constraints from the longitudinal polarization fraction of the $D^*$ meson, $F_L$,  and another angular observable $\tilde{F}_L$, which are proportional to the coefficients  of the $\cos^2 \theta^*$ and $\cos^2 \theta_\ell$ terms in the angular distribution, respectively. These quantities were extracted for the first time by \cite{Bobeth:2021lya} using the binned CP-averaged differential decay distribution data provided by Belle \cite{Belle:2019gij}. They obtain a CP-averaged SM prediction for the integrated $\langle \Delta F_L \rangle$ and $\langle \Delta \tilde{F}_L \rangle$ to be $(5.43 \pm 0.36) \times 10^{-4}$ and $(-5.20\pm 0.30)\times 10^{-3}$ respectively. By fitting the data, they also report $\langle \Delta F_L \rangle^{exp} = -0.0065 \pm 0.0059$ and $\langle \Delta \tilde{F}_L \rangle^{exp} = -0.0107 \pm 0.0142$. We have verified that our benchmark values satisfy these experimental bounds within a $1\sigma$ confidence interval.

\section{Conclusions} \label{sec:conc}

Motivated by the $\Delta A_{FB}$ anomaly in $\oB\to D^{*+}\mu^- \bar{\nu}$ decay, which could be a sign of physics beyond the Standard Model \cite{Bobeth:2021lya}, we have developed a new Monte Carlo New Physics (NP) generator tool for $B \to D^*\ell\nu_\ell$ with $\ell= e, \mu, \tau$ in the EvtGen framework \cite{Campagna:2022evt}. The full theoretical description for the effective basis we use to parameterize NP as well as the different angular asymmetries has been comprehensively discussed in this article. We used this tool to examine signatures of NP, which are consistent with current data and with the hints of NP in $B \to D^*\mu \nu_\mu$ assuming that the decay $B \to D^* e \nu_e$ is well described by the SM. We found that the angular asymmetries, $A_{FB}, S_5, S_3,$ and $S_7$, which can be extracted from the fully reconstructed angular distribution, are sensitive to new physics. With current experimental constraints, we show the part of the $g_i$ NP parameter space that is still allowed (see Fig.~\ref{fig:param_plot}).

We introduce the $\Delta$ observables, which are obtained by taking the differences between the observables for the muon and the electron modes, in order to avoid theory uncertainties due to form factors, which might obscure signals of NP. We suggest experimental requirements on $q^2$ and lepton momenta in order to increase sensitivity to NP and reduce systematics. We identify $\Delta A_{FB}$ and $\Delta S_5$ as the most powerful probes of NP with little sensitivity to form-factor uncertainties; this is shown in Fig.~\ref{fig:CLN_HQET_BGL}.
We also observe that correlated signatures of NP in multiple observables such as $\Delta A_{FB}$ and $\Delta S_5$ are required to confirm the presence of NP (see Fig.~\ref{fig:observables}.) Therefore, if a NP signal for $\Delta A_{FB}$ is observed in future experiments, it must be accompanied by a corresponding signal in $\Delta S_5$ both in the integrated variable and the $q^2$ distribution. We calculate integrated observables and plot coarse binned expectations for $\Delta A_{FB}$ and $\Delta S_5$, as well as correlations between the two. The NP signatures described here are ideally suited for Belle II at 1, 5, 50, and 250 ab$^{-1}$ and might also be explored at hadron collider experiments.

{\textbf{Note Added :}} This paper is an improved version of Ref.~\cite{Bhattacharya:2022cna}, which was submitted to the US Community Summer Study on the Future of Particle Physics (Snowmass 2021) but will not appear in the final proceedings. Improvements include calculations of correlations between several observables and discussions of prospects for NP-sensitive observables with several benchmark values of Belle II integrated luminosity.

\begin{acknowledgments}
This work was supported in part by the National Science Foundation under Grant No.~PHY-2013984 (B.B.) and PHY-1915142 (Q.C., A.D., and L.M.). T.E.B, S.D., and A.S.\ acknowledge support from the (DOE) Office of High Energy Physics (OHEP) Award No. DE-SC0010504. The work of B.B. was completed with partial support from the Munich Institute for Astro- and Particle Physics (MIAPP) which is funded by the Deutsche Forschungsgemeinschaft (DFG, German Research Foundation) under Germany's Excellence Strategy – EXC-2094 – 390783311. B.B. additionally thanks D.~Van Dyk for useful conversations. L.M. thanks Honkai Liu for fruitful discussions regarding HQET form factors.
\end{acknowledgments}

\appendix\label{ap:A}

\section{Angular coefficients}\label{sec:iiscs}

The angular distribution of $\oB\to D^*\ell^-{\bar\nu}$ presented in Eq.~(\ref{eq:angdist}) contains 12 coefficients labeled $I^{(s,c)}_i$ with $i = 1,\ldots,9$. The full list of angular coefficients are presented below as functions of eight helicity  amplitudes, $\cA_{SP},\cA_t,\cA_0,\cA_{||},\cA_\perp,\cA_{0T},\cA_{||,T},$ and $\cA_{\perp,T}$. These helicity amplitudes depend on hadronic form factors as well as NP coefficients. The form of the eight helicity amplitudes are given in Appendix \ref{sec:ffs}.
\bea
I_i^{(s,c)} &=& \frac{G_F^2|V_{cb}|^2(q^2 - m_\ell^2)^2|p_{D^*}|}{192\pi^3m^2_Bq^2}\cB(D^*\to D\pi) {\tilde I}_i^{(s,c)}, \\
{\tilde I}_1^c &=& 4\,\(|\cA_{SP}|^2 + \frac{m^2_\ell}{q^2}|\cA_t|^2\) + 2\,\(1 + \frac{m^2_\ell}{q^2}\)\(|\cA_0|^2 + 16\,|\cA_{0,T}|^2\) ~\nl
&&\hspace{5truemm}+~8\,\frac{m_\ell}{\sqrt{q^2}}\lb{\rm Re}\[\cA_t\cA_{SP}^*\] - 4\,{\rm Re}\[\cA_0\cA_{0,T}^*\]\rb ,~ \\
{\tilde I}_1^s &=& \lb\frac{3}{2}(|\cA_{||}|^2 + |\cA_{\perp}|^2) + 8(|\cA_{||,T}|^2 + |\cA_{\perp,T}|^2)\rb ~\nl
&&\hspace{5truemm}-~16\,\frac{m_\ell}{\sqrt{q^2}}\lb{\rm Re}[\cA_{||}\cA_{||,T}^*] + {\rm Re}[\cA_{\perp}\cA_{\perp,T}^*]\rb ~\nl
&&\hspace{5truemm}+~\frac{m_\ell^2}{q^2}\lb\frac{1}{2}\(|\cA_{||}|^2 + |\cA_\perp|^2\) + 24\(|\cA_{||,T}|^2 + |\cA_{\perp,T}|^2\)\rb,~\\
{\tilde I}_2^c &=& -2\(1 - \frac{m^2_\ell}{q^2}\)\lb|\cA_0|^2 - 16|\cA_{0,T}|^2\rb,~\\
{\tilde I}_2^s &=& \frac{1}{2}\(1 - \frac{m^2_\ell}{q^2}\)\lb\(|\cA_{||}|^2 + |\cA_{\perp}|^2\) - 16\(|\cA_{||,T}|^2 + |\cA_{\perp,T}|^2\)\rb,~\\
{\tilde I}_3 &=& -\(1 - \frac{m^2_\ell}{q^2}\)\lb\(|\cA_{||}|^2 - |\cA_\perp|^2\) - 16\(|\cA_{||,T}|^2 - |\cA_{\perp,T}|^2\)\rb,~\\
{\tilde I}_4 &=& \sqrt{2}\(1 - \frac{m^2_\ell}{q^2}\)\lb16\,{\rm Re}\[\cA_{0,T}\cA_{||,T}^*\] - {\rm Re}\[\cA_0\cA_{||}^*\]\rb,~\\
{\tilde I}_5 &=& 2\sqrt{2}\,\lb\({\rm Re}\[\cA_0\cA_\perp^*\] + 4\,{\rm Re}\[\cA_{||,T}\cA^*_{SP}\]\) + \frac{m_\ell^2}{q^2}\(16\,{\rm Re}\[\cA_{0,T}\cA_{\perp,T}^*\] - {\rm Re}\[\cA_{||}\cA_t^*\]\) \r. \nl
&&\hspace{1truecm}\l.+~\frac{m_\ell}{\sqrt{q^2}}\(4\,{\rm Re}\[\cA_{||,T}\cA_t^*\] - 4\,{\rm Re}\[\cA_0\cA_{\perp,T}^*\] - 4\,{\rm Re} \[\cA_{0,T}\cA_\perp^*\] - \,{\rm Re}\[\cA_{||}\cA_{SP}^*\]\)\rb,~\\
{\tilde I}_6^c &=& 32\,{\rm Re}\[\cA_{0,T}\cA_{SP}^*\] + \frac{m_\ell}{\sqrt{q^2}}\,\lb32\,{\rm Re}\[\cA_{0,T}\cA_t^*\] - 8 {\rm Re}\[\cA_0\cA_{SP}^*\]\rb -8\,\frac{m^2_\ell}{q^2}{\rm Re}\[\cA_0\cA_t^*\],~\\
{\tilde I}_6^s &=& -~4\,{\rm Re}\[\cA_{||}\cA_{\perp}^*\] + 16\,\frac{m_\ell}{\sqrt{q^2}}\,\lb{\rm Re}\[\cA_{||}\cA^*_{\perp,T}\] + {\rm Re}\[\cA_{||,T}\cA_\perp^*\]\rb - 64\,\frac{m^2_\ell}{q^2}\,{\rm Re}\[\cA_{||,T}\cA_{\perp,T}^*\],~\\
{\tilde I}_7 &=& -~8\sqrt{2}\,{\rm Im}\[\cA_{SP}\cA_{\perp,T}^*\] - 2\sqrt{2}\,{\rm Im}\[\cA_0\cA_{||}^*\] + 2\sqrt{2}\,\frac{m^2_\ell}{q^2}\,{\rm Im}\[\cA_t\cA^*_\perp\]~ \nl
&& \hspace{5truemm} +~2\sqrt{2}\,\frac{m_\ell}{\sqrt{q^2}}\lb4\,{\rm Im}\[\cA_0\cA_{||,T}^*\] - 4\,{\rm Im}\[\cA_{||}\cA_{0,T}^*\]  - 4\,{\rm Im}\[\cA_t\cA_{\perp,T}^*\] -  {\rm Im}\[\cA_\perp \cA_{SP}^*\]\rb,~\\
{\tilde I}_8 &=& -\sqrt{2}\,\(1 - \frac{m^2_\ell}{q^2}\){\rm Im}\[\cA_{\perp} \cA_0^*\],~ \\
{\tilde I}_9 &=& 2\,\(1 - \frac{m^2_\ell}{q^2}\){\rm Im}\[\cA_{||} \cA_\perp^*\],~
\eea
where $|p_{D^*}| = \sqrt{\lambda(m^2_B,m^2_{D^*},q^2)}/(2m_B)$ represents the magnitude of the $D^*$ 3-momentum, and $\lambda(a,b,c) = a^2 + b^2 + c^2 - 2 a b - 2 b c - 2 c a$.

\section{Helicity Amplitudes and Form Factors}\label{sec:ffs}

The 12 angular coefficients needed to construct the full angular distribution of Eq.~(\ref{eq:angdist}) were presented in Appendix \ref{sec:iiscs}. These angular coefficients depend on eight helicity amplitudes that can be further expressed in terms of NP coefficients ($g_P, g_L, g_R$, and $g_T$) and hadronic form factors. We list the helicity amplitudes below \cite{Sakaki:2013bfa,Beneke:2000wa}.
\bea
\cA_{SP} &=& -g_P\,\frac{\sqrt{\lambda(m_B^2,\mDs^2, q^2)}}{m_b+m_c} A_0(q^2) ,~ \\
\cA_{0} &=& -\frac{(1+g_L-g_R)(m_B + \mDs)}{2\mDs\sqrt{q^2}}\[(m_B^2-\mDs^2 - q^2) A_1(q^2) - \frac{\lambda(m_B^2, \mDs^2 , q^2)}{(m_B + \mDs)^2} A_2(q^2)\],~ \\
\cA_t &=& -(1+g_L-g_R) \, \frac{\sqrt{\lambda(m_B^2,\mDs^2, q^2)}}{\sqrt{q^2}} A_0(q^2) ,~\\
\mathcal{A}_{\pm} &=& (1+g_L-g_R) \, (m_B + \mDs)A_1(q^2) \mp (1+g_L+g_R)\frac{\sqrt{\lambda(m_B^2,\mDs^2, q^2)}}{m_B + \mDs} V(q^2),~ \\
\cA_{0,T} &=& \frac{g_T}{2\mDs(m_B^2-\mDs^2)} \( (m_B^2-\mDs^2)(m_B^2+3\mDs^2 -q^2)T_2(q^2) -\lambda(m_B^2,\mDs^2, q^2)T_3(q^2) \) ,~\\
\cA_{\pm, T} &=&  g_{T} \, \frac{\sqrt{\lambda(m_B^2,\mDs^2, q^2)}T_1(q^2) \pm (m_B^2-\mDs^2)T_2(q^2)}{\sqrt{q^2}} ,~
\eea
The angular coefficients requiring vector and/or tensor type contributions may also require the amplitudes to be expressed in the transversity basis as follows.
\bea
\cA_{||,T} &=& \(\cA_{+(,T)} + \cA_{-(,T)}\)/\sqrt{2} ,~ \\
\cA_{\perp,T} &=& \(\cA_{+(,T)} - \cA_{-(,T)}\)/\sqrt{2} .~
\eea

The above helicity amplitudes depend on the seven hadronic form factors listed below.
\bea
      V(q^2) &=& \frac{m_B + m_\Dst}{2\sqrt{m_B m_\Dst}} h_V(w(q^2)) , \label{eq:FF-HQET-relation0}\\
      A_1(q^2) &=& \frac{( m_B + m_\Dst )^2 - q^2}{2\sqrt{m_B m_\Dst} ( m_B + m_\Dst)} h_{A_1}(w(q^2)) , \\
      A_2(q^2) &=& \frac{m_B+m_\Dst}{2\sqrt{m_B m_\Dst}} \left[ h_{A_3}(w(q^2)) + \frac{m_\Dst}{m_B} h_{A_2}(w(q^2)) \right], \\
      A_0(q^2) &=& \frac{1}{2\sqrt{m_B m_\Dst}} \left[\frac{(m_B + m_\Dst)^2 - q^2}{2m_\Dst} h_{A_1}(w(q^2)) \right. \nn  \\
                && \hspace{2truecm} -\left. \frac{m_B^2 - m_\Dst^2 + q^2}{2m_B} h_{A_2}(w(q^2)) - \frac{m_B^2 - m_\Dst^2 - q^2}{2m_\Dst} h_{A_3}(w(q^2)) \right], \\
      T_1(q^2) &=& \frac{1}{2\sqrt{m_B m_\Dst}} \left[(m_B + m_\Dst) h_{T_1}(w(q^2)) - (m_B - m_\Dst ) h_{T_2}(w(q^2)) \right], \\
      T_2(q^2) &=& \frac{1}{2\sqrt{m_B m_\Dst}} \left[\frac{( m_B + m_\Dst)^2 - q^2}{m_B + m_\Dst} h_{T_1}(w(q^2)) \right. \nn \\
                && \hspace{2truecm} \left. - \frac{(m_B - m_\Dst )^2 - q^2}{m_B - m_\Dst} h_{T_2}(w(q^2)) \right], \\
      T_3(q^2) &=& \frac{1}{2\sqrt{m_B m_\Dst}} \left[(m_B - m_\Dst) h_{T_1}(w(q^2)) - (m_B + m_\Dst) h_{T_2}(w(q^2)) \right. \nn \\
                && \hspace{2truecm} \left.- 2 \frac{m_B^2 - m_\Dst^2} {m_B} h_{T_3}(w(q^2)) \right],
    \label{eq:FF-HQET-relation}
\eea
where the recoil angle, $w(q^2)$ can be expressed as is $w(q^2)=(m_B^2+m_{D^{*}}^2-q^2)/2m_Bm_{D^{*}}$. The above expressions depend on several lepton and meson masses that are used as input parameters. In our calculations we use the values of meson and lepton masses given in Table \ref{tab:inputs-mass}.
\begin{table}[h]
\begin{tabular}{|c|c|}
\hline
Masses & Value (MeV) \\
\hline
$m_{B^0}$ & $5279.63(20)$ \\
$m_{D^{*+}}$ & $2010.26(05)$ \\
$m_e$ & $0.5109989461(31)$ \\
$m_\mu$ & $105.6583745(24)$ \\ \hline
\end{tabular}
\caption{Input values used for meson and lepton masses taken from the Particle Data Group \cite{Zyla:2020zbs}. Numbers in parentheses represent the errors in the last two digits.}
\label{tab:inputs-mass}
\end{table}
We have also used the following values for the quark masses, $m_b = $ GeV and $m_c = $ GeV.

Note that the above form factors still depend on several additional functions of $q^2$, namely $h_V$, $h_{A_1}$, $h_{A_2}$, $h_{A_3}$, $h_{T_1}$, $h_{T_2}$, $h_{T_3}$, $R_1$, $R_2$, and $R_3$. There are several ways of parameterizing these functions using Heavy Quark Effective Theory (HQET). Two such parameterizations are presented in Appendix \ref{sec:ffps}.  

\section{Parameterizations of the hadronic form factors} \label{sec:ffps}

The hadronic form factors described in Appendix \ref{sec:ffs} depend on several form factors that appear as functions of $q^2$ in HQET. At present there are several ways of parameterizing these functions. Although each parameterization gives a slightly different value for the underlying function, a conclusive identification of the best way to parameterize these functions still eludes us. This problem adds to the theoretical uncertainties associated with the determinations of some of the NP observables discussed in this article. 

A commonly used parameterization for the HQET form factors, first presented by Caprini, Lellouch, and Neubert (CLN) in Ref.~\cite{Caprini:1997mu} is given below.
\bea
    h_V(w) &=& R_1(w) h_{A_1}(w), \\
    h_{A_2}(w) &=& \frac{R_2(w)-R_3(w)}{2 r_\Dst } h_{A_1}(w), \\
    h_{A_3}(w) &=& \frac{R_2(w)+R_3(w)}{2} h_{A_1}(w), \\
    h_{T_1}(w) &=& \frac{1}{2 ( 1 + r_\Dst^2 - 2r_\Dst w ) } \left[ \frac{m_b - m_c}{m_B - m_\Dst} ( 1 - r_\Dst)^2 ( w + 1 ) \, h_{A_1}(w) \right. \nn\\
      & & \quad\quad\quad\quad\quad\quad\quad\quad\quad \left. - \frac{m_b + m_c}{m_B + m_\Dst} ( 1 + r_\Dst )^2 ( w - 1 ) \, h_V(w) \right] \,, \\
    h_{T_2}(w) &=& \frac{( 1 - r_\Dst^2)(w + 1)}{2( 1 + r_\Dst^2 - 2r_\Dst w)} \left[\frac{m_b - m_c}{m_B - m_\Dst} \, h_{A_1}(w) - \frac{m_b + m_c}{m_B + m_\Dst} \, h_V(w) \right] \,, \\
    h_{T_3}(w) &=& -\frac{1}{2 ( 1 + r_\Dst ) ( 1 + r_\Dst^2 - 2r_\Dst w ) } \left[2 \frac{m_b - m_c}{m_B - m_\Dst } r_\Dst ( w + 1 ) \, h_{A_1}(w) \right. \nn\\
      & & + \frac{m_b - m_c}{m_B - m_\Dst } ( 1 + r_\Dst^2 - 2r_\Dst w ) ( h_{A_3}(w) - r_\Dst h_{A_2}(w) ) \nn\\
      & & \left. - \frac{m_b + m_c}{m_B + m_\Dst} ( 1 + r_\Dst )^2 \, h_V(w) \right] \,,
    \label{eq:HQET-CLN}
\eea
where $r_{D^{*}} = m_{D^{*}}/m_B$ and the $w$-dependencies are
expressed as
\bea
      h_{A_1}(w) &=& h_{A_1}(1) \left[ 1 - 8\rho_\Dst^2 z + (53\rho_\Dst^2-15) z^2 - (231\rho_\Dst^2-91) z^3 \right] \\
      R_1(w) &=& R_1(1) - 0.12(w-1) + 0.05(w-1)^2, \\
      R_2(w) &=& R_2(1) + 0.11(w-1) - 0.06(w-1)^2, \\
      R_3(w) &=& 1.22 - 0.052(w-1) + 0.026(w-1)^2.
      \label{eq:HQET_w_parametrization}
\eea
The parameter $z$ is related to the recoil angle $w$ through $z(w) = (\sqrt{w+1} - \sqrt2 ) / ( \sqrt{w+1} + \sqrt2 )$. The values of $h_{A_1}(1)$, $R_1(1)$, $R_2(1)$, and $\rho_{D^*}^2$, listed in Table~\ref{tab:inputs-others}, were taken from Ref.~\cite{Sakaki:2013bfa}.
\begin{table}[h]
\begin{tabular}{|c|c|}
\hline
Parameter & Value \\
\hline
$h_{A_{1}}(1)$ & $0.908\pm 0.017$\\
$\rho_{D^*}^2$ & $1.207\pm 0.026$\\
$R_1(1)$ & $1.403\pm 0.033$\\
$R_2(1)$ & $0.854\pm 0.020$\\
\hline
\end{tabular}
\caption{Input values of parameters needed for the CLN parameterization of form factors used here were taken from \cite{Sakaki:2013bfa}.}
\label{tab:inputs-others}
\end{table}

Yet another way of parameterizing the HQET form factors is to express them in terms of the leading Isgur-Wise (IW) function $\xi (w)$ \cite{Isgur:1990jf} and sub-leading IW terms, which represents higher order power corrections to the leading IW function as 
\beq
h_X (w) = \xi(w) \hh_X (w), ~~~(X = V, A_1, A_2, A_3, T_1, T_2, T_3)
\eeq
where 
\beq
\hh_X (w) = \hh_{X,0} + \varepsilon_a~ \delta\hh_{X,\as} + \varepsilon_b~ \delta\hh_{X,m_b} + \varepsilon_c~ \delta\hh_{X,m_c} + \varepsilon_c^2~ \delta\hh_{X,m_c^2}.
\label{eq:hh}
\eeq
Here, $\varepsilon_a, \varepsilon_b, \varepsilon_c$ denote the expansion coefficients corresponding to the higher order corrections in $\as$ and $1/m_{b,c}$ respectively which were worked out by \cite{Neubert:1993mb, Caprini:1997mu} using heavy quark symmetry.

\begin{table}[h]
\begin{tabular}{|c|c|c|}
\hline
Parameter & HQET (3/2/1) & HQET (2/1/0) \\
\hline
$\xi^{(0)}$ & $1$ & $1$\\
$\xi^{(1)}$ & $-0.93 \pm 0.10$ & $-1.10 \pm 0.04$\\
$\xi^{(2)}$ & $+1.35 \pm 0.26$ & $+1.57 \pm 0.10$\\
$\xi^{(3)}$ & $-2.67 \pm 0.75$ & $-$\\
\hline
\hline
$\hat{\chi}_2^{(0)}$ & $-0.05 \pm 0.02$ & $-0.06 \pm 0.02$\\
$\hat{\chi}_2^{(1)}$ & $+0.01 \pm 0.02$ & $-0.06 \pm 0.02$\\
$\hat{\chi}_2^{(2)}$ & $-0.01 \pm 0.02$ & $-$\\
$\hat{\chi}_3^{(0)}$ & $0$ & $0$\\
$\hat{\chi}_3^{(1)}$ & $-0.05 \pm 0.02$ & $-0.03 \pm 0.01$\\
$\hat{\chi}_3^{(2)}$ & $-0.03 \pm 0.03$ & $-$\\
\hline
\hline
$\eta^{(0)}$ & $+0.74 \pm 0.11$ & $+0.38 \pm 0.06$\\
$\eta^{(1)}$ & $+0.05 \pm 0.03$ & $+0.08 \pm 0.03$\\
$\eta^{(2)}$ & $-0.05 \pm 0.05$ & $-$\\
\hline
\hline
$\tilde{\ell}_1^{(0)}$ & $+0.09 \pm 0.18$& $+0.50 \pm 0.16$\\
$\tilde{\ell}_1^{(1)}$ & $+1.20 \pm 2.09$ & $-$\\
$\tilde{\ell}_2^{(0)}$ & $-2.29 \pm 0.33$ & $-2.16 \pm 0.29$\\
$\tilde{\ell}_2^{(1)}$ & $-3.66 \pm 1.56$ & $-$\\
$\tilde{\ell}_3^{(0)}$ & $-1.90 \pm 12.4$ & $-1.14 \pm 2.34$\\
$\tilde{\ell}_3^{(1)}$ & $+3.91 \pm 4.35$ & $-$\\
$\tilde{\ell}_4^{(0)}$ & $-2.56 \pm 0.94$ & $+0.82 \pm 0.47$\\
$\tilde{\ell}_4^{(1)}$ & $+1.78 \pm 0.93$ & $-$\\
$\tilde{\ell}_5^{(0)}$ & $+3.96 \pm 1.17$ & $+1.39 \pm 0.43$\\
$\tilde{\ell}_5^{(1)}$ & $+2.10 \pm 1.47$ & $-$\\
$\tilde{\ell}_6^{(0)}$ & $+4.96 \pm 5.76$ & $+0.17 \pm 1.15$\\
$\tilde{\ell}_6^{(1)}$ & $+5.08 \pm 2.97$ & $-$\\
\hline
\end{tabular}
\caption{Values of input parameters needed for the HQET (3/2/1) and HQET (2/1/0) parameterizations of the hadronic form factors taken from \cite{Iguro:2020cpg}.}
\label{tab:FFparam}
\end{table}

The leading term in \ref{eq:hh} is
\beq
\hh_{X,0} = \begin{cases}
1~~\rm{for}~~X = A_1, A_3, T_1, \\
0~~\rm{for}~~X = A_2, T_2, T_3.
\end{cases}
\eeq 
The $\as$ corrections are given as 
\begin{align}
 \dhh_{V,\as} &= 
 \frac{1}{6\zcb (w - \wcb)} 
 \left[ 4\zcb (w-\wcb) \Omega_w (w) + 2(w+1)((3w-1)\zcb -\zcb^2-1) r_w (w) \right. \nn \\
 &\, \quad\quad\quad\quad\quad\quad\quad\quad \left. -12\zcb (w-\wcb) -(\zcb^2-1) \log \zcb \right] + V(\mu) \,, \\
 \dhh_{A_1,\as} &= 
 \frac{1}{6\zcb (w - \wcb)} 
 \left[ 4\zcb (w-\wcb) \Omega_w (w) + 2(w-1)((3w+1)\zcb -\zcb^2-1) r_w (w) \right. \nn \\
 &\, \quad\quad\quad\quad\quad\quad\quad\quad \left. -12\zcb (w-\wcb) -(\zcb^2-1) \log \zcb \right] + V(\mu) \,,\\
\dhh_{A_2,\as} &= 
 \frac{-1}{6\zcb^2 (w - \wcb)^2} 
\left[ \left(2 + (2w^2-5w-1)\zcb  +2w(2w-1)\zcb^2  + (1-w)\zcb^3 \right) r_w (w) \right. \nl 
&\, \quad\quad\quad\quad\quad\quad\quad\quad \left.
 -2\zcb(\zcb+1)(w-\wcb) + (\zcb^2-(4w+2)\zcb +3 +2w) \zcb \log \zcb \right] \,,\\
\dhh_{A_3,\as} &= \dhh_{A_1,\alpha_s} + \frac{1}{6\zcb (w - \wcb)^2} \[ \,2 \zcb (\zcb +1) (\wcb - w) + \big( 2\zcb^3 + \zcb^2 (2w^2-5w-1) \right. \nl
&\, \quad\quad\quad\quad\quad\quad\quad\quad\quad\quad\quad\quad \left. + \zcb (4w^2-2w) - w+1 \big) r_w (w) -\big( \zcb^2 (2w+3) \right. \nn \\ 
&\, \quad\quad\quad\quad\quad\quad\quad\quad\quad\quad\quad\quad \left. - \zcb (4w+2) +1 \big) \log \zcb \] \\
\dhh_{T_1,\as} &= 
 \frac{1}{3\zcb (w - \wcb)} 
\left[ 2\zcb (w-\wcb) \Omega_w (w) + 2\zcb (w^2-1) r_w (w)  -6\zcb (w-\wcb) \right. \nl 
&\, \quad\quad\quad\quad\quad\quad\quad\quad \left. +(1-\zcb^2) \log \zcb \right] +T(\mu) \,, \\
 \dhh_{T_2,\as} &= 
 \frac{w+1}{3\zcb (w - \wcb)} 
 \left[ (1-\zcb^2) r_w (w) +2 \zcb \log \zcb \right] \,, \\
 \dhh_{T_3,\as} &= 
 \frac{1}{3\zcb (w - \wcb)} 
 \left[ (\zcb w-1) r_w (w) - \zcb \log \zcb \right] \, ,
\end{align}
where  
\begin{align}
 & \zcb = \frac{m_c}{m_b} \,, \quad \wcb = \frac{1}{2} \left( \zcb + \zcb^{-1} \right) \,, \quad w_\pm(w) = w \pm \sqrt{w^2-1} \,, \\
 & r_w (w) = \frac{\log w_+ (w)}{\sqrt{w^2-1} }\,, \\
 & \Omega_w (w) = \frac{w}{2\sqrt{w^2-1}} \Big[2\text{Li}_2 (1-w_-(w)\zcb) - 2\text{Li}_2 (1-w_+(w)\zcb) \notag \\
 & \hspace{9em}+ \text{Li}_2 (1-w_+^2(w)) - \text{Li}_2 (1-w_-^2(w)) \Big] - w r_w(w) \log \zcb + 1 \,.
\end{align}
Here $\text{Li}_2(x) = \int\limits_x^0 dt \log(1-t)/t$ is the dilogarithm function and $V(\mu),T(\mu)$ are scale factors given as
\bea
V(\mu) &=& -\frac{2}{3} \big( wr_w(w)-1 \big) \log \frac{m_bm_c}{\mu^2} \,, \\
T(\mu) &=& -\frac{1}{3} \big( 2wr_w(w)-3 \big) \log \frac{m_bm_c}{\mu^2} \,.
\eea 
In our calculations we choose the scale $\mu = 4.2$ GeV.
The $1/m_{b,c}$ corrections in eq.~\ref{eq:hh} are given as 
\bea
 \dhh_{V,m_b} 
 &=& \hL_1(w) - \hL_4(w) \,, \\
 \dhh_{V,m_c} &=&
 \hL_2(w) - \hL_5(w) \,, \\
 \dhh_{A_1,m_b} 
&=& \hL_1(w) - \frac{w-1}{w+1}\hL_4(w)  \,, \\
 \dhh_{A_1,m_c} 
 &=& \hL_2 - \frac{w-1}{w+1}\hL_5(w)  \,, \\
 \dhh_{A_2,m_b} &=& 0\,, \\
 \dhh_{A_2,m_c} &=&
 \hL_3(w) + \hL_6(w)  \,, \\
 \dhh_{A_3,m_b} &=&
\hL_1(w) - \hL_4(w) \,, \\
 \dhh_{A_3,m_c} &=& 
  \hL_2(w) - \hL_3(w) + \hL_6 (w)- \hL_5(w) \,, \\
 \dhh_{T_1,m_b} &=& \hL_1(w) \,, \\
 \dhh_{T_1,m_c} &=& \hL_2(w) \,, \\
 \dhh_{T_2,m_b} &=&  -\hL_4(w) \,, \\
 \dhh_{T_2,m_c} &=& \hL_5(w) \,, \\
 \dhh_{T_3,m_b} &=& 0 \,, \\
 \dhh_{T_3,m_c} &=& \frac{1}{2}\left(\hL_6(w) - \hL_3(w)\right) \,, 
\eea
where the $\hL(w)$ functions read
\bea
\hL_1(w) &=& -4(w-1) \hat\chi_2(w) + 12 \hat\chi_3(w) \,,\\
\hL_2(w) &=& -4\hat\chi_3(w) \,,\\
\hL_3(w) &=& 4\hat\chi_2(w)\,,\\
\hL_4(w) &=& 2\eta (w) -1 \,,\\
\hL_5(w) &=& -1 \,, \\
\hL_6(w) &=& -\frac{2 (1+\eta(w))}{w+1}
\label{eq:Lhat}
\eea

The corrections of order $1/m_c^2$ are included via the subleading reduced IW functions $\hat l_{1-6}(w)$ as \cite{Falk:1992wt,Bordone:2019vic}
\begin{align}
 \dhh_{V,m_c^2} & = \hl_2 (w) - \hl_5 (w) \,, \\
 \dhh_{A_1,m_c^2} & = \hl_2 (w) - \frac{w-1} {w+1} \hl_5 (w) \,, \\
 \dhh_{A_2,m_c^2} & = \hl_3 (w) + \hl_6 (w) \,, \\
 \dhh_{A_3,m_c^2} & = \hl_2 (w) - \hl_3 (w) - \hl_5 (w) + \hl_6 (w)  \,, \\
 \dhh_{T_1,m_c^2} & = \hl_2 (w) \,, \\
 \dhh_{T_2,m_c^2} & = \hl_5 (w) \,, \\
 \dhh_{T_3,m_c^2} & = \frac{1}{2} \big( \hl_3 (w) -\hl_6 (w) \big) \,.
\end{align}

The IW functions are expressed, in general, as expansions about $w=1$ as
\beq
f(w) = \sum_{n=0} \frac{f^{(n)}}{n!}(w-1)^n
\eeq
with $f = \xi, \eta, \hat{\chi}_2, \hat{\chi}_3$ and $\hl_i$. One can further relate the kinematic variable $w$ with the expansion variable $z$ as
\beq
w(z) = 2\left(\frac{1+z}{1-z}\right)^2 -1.
\label{eq:z-w-relation}
\eeq
One can then expand the IW functions up to any order in z as
\beq 
f(w) = f^{(0)} + 8 f^{(1)} z + 16 (f^{(1)} + 2 f^{(2)}) z^2 + \frac{8}{3}(9 f^{(1)} + 48 f^{(2)} + 32 f^{(3)})z^3 + .... (\rm{higher~orders}).
\eeq 
The authors of Ref.~\cite{Iguro:2020cpg} have performed a simultaneous fit of the HQET parameters and the CKM element $V_{cb}$ by considering an expansion of the IW functions up to order NNLO (3/2/1) and NNLO (2/1/0) where 
\bea 
{\rm NNLO}~(3/2/1) &:& \xi(w) \rm{~up~to~} z^3, \hat\chi_{2,3} (w), \eta (w) \rm{~up~to~order~} z^2 \rm{~and~} \hl_i \rm{~up~to~order~} z \\
{\rm NNLO}~(2/1/0) &:& \xi(w) \rm{~up~to~} z^2, \hat\chi_{2,3} (w), \eta (w) \rm{~up~to~order~} z \rm{~and~} \hl_i \rm{~up~to~order~} z^0.
\eea
The fitted value of the parameters for the above two scenarios from Ref.~\cite{Iguro:2020cpg} are given in Table~\ref{tab:FFparam}.

\begin{table}[t]
\begin{tabular}{|c|c|c|}
\hline
Form Factor & Type & Pole Masses $M_p$ (GeV) \\
\hline
$g$ & $1^-$ & $6.329, 6.920, 7.020$\\
$f, \mathcal{F}_1$ & $1^+$ & $6.739, 6.750, 7.145, 7.150$\\
$\mathcal{F}_2$ & $0^-$ & $6.275, 6.842, 7.250$ \\
\hline
\end{tabular}
\caption{The pole masses corresponding to different types of $B_c$ resonances as listed in \cite{Bigi:2017jbd}.}
\label{tab:BGL-PoleMasses}
\end{table}

The other alternate way of parameterizing the form factors is due to Boyd, Grinstein and Lebed (BGL) \cite{Boyd:1997kz}. Both the CLN and BGL 
 form factor coefficients are constrained from the same dispersive bounds. However, unlike CLN, they do not employ HQET relations to reduce the number of form factor parameters and are hence, more general. The form factors $\mathcal{F}_i \equiv \{f,g,\mathcal{F}_1,\mathcal{F}_2\}$ are expressed as series expansions in $z$ as
\beq
\mathcal{F}_i(z) = \frac{1}{P_i(z) \phi_i(z)} \sum_{j=0}^N a_j^{\mathcal{F}_i} z^j,
\eeq 
where $z$ is related to the recoil angle $w$ as in Eq.~\eqref{eq:z-w-relation} and $P_i(z) = \prod_p \frac{z-z_p}{1-z z_p}$ are called the Blaschke factors that help eliminate poles at $z=z_p$ at the $B_c$ resonances given by
\beq
z_p = \frac{\sqrt{t_+ -M_p^2}-\sqrt{t_+ -t_-}}{\sqrt{t_+ -M_p^2}+\sqrt{t_+ -t_-}};~~~t_\pm = (m_B \pm m_{D^*})^2.
\eeq 
The pole mass ($M_p$) for the different types of resonances are listed in Table~\ref{tab:BGL-PoleMasses}. The outer functions $\phi_i$ are given as
\bea 
\phi_f &=& \frac{4r_{D^{*}}}{m_B^2} \sqrt{\frac{n_I}{6\pi \chi_{1^+}^T (0)}} \frac{(1+z)(1-z)^{3/2}}{\left[(1+r_{D^{*}})(1-z) + 2\sqrt{r_{D^{*}}}(1+z)\right]^4}, \\
\phi_g &=& 16r_{D^{*}}^2 \sqrt{\frac{n_I}{3\pi \tilde{\chi}_{1^-}^T (0)}} \frac{(1+z)^2(1-z)^{-1/2}}{\left[(1+r_{D^{*}})(1-z) + 2\sqrt{r}(1+z)\right]^4},\\
\phi_{\mathcal{F}_1} &=& \frac{4r_{D^{*}}}{m_B^3} \sqrt{\frac{n_I}{6\pi \chi_{1^+}^T (0)}} \frac{(1+z)(1-z)^{5/2}}{\left[(1+r_{D^{*}})(1-z) + 2\sqrt{r_{D^{*}}}(1+z)\right]^5},  \\
\phi_{\mathcal{F}_2} &=& 8\sqrt{2}r_{D^{*}}^2 \sqrt{\frac{n_I}{\pi \tilde{\chi}_{1^+}^L (0)}} \frac{(1+z)^2 (1-z)^{-1/2}}{\left[(1+r_{D^{*}})(1-z) + 2\sqrt{r_{D^{*}}}(1+z)\right]^4}.
\label{eq:BGL-outer-func}
\eea 
The various relevant inputs for computing the outer functions are listed in Table~\ref{tab:BGL-other-inputs}. The form factor coefficients $a_j^{\mathcal{F}_i}$ satisfy the weak unitarity constraints given by
\beq
\sum_{j=0}^{N} (a_j^{g})^2 < 1,~~\sum_{j=0}^{N} (a_j^{f})^2 + (a_j^{\mathcal{F}_1})^2  < 1, ~~\sum_{j=0}^{N} (a_j^{\mathcal{F}_2})^2  < 1.
\label{eq:FF-unity-constr}
\eeq
In addition to this, they are also subject to two kinematic constraints, one each at zero and maximum recoil respectively, given by :
\bea
\mathcal{F}_1 (1) &=& m_B (1-r_{D^{*}}) f(1),\\
\mathcal{F}_2 (w_{max}) &=& \frac{1+r_{D^{*}}}{m_B^2 (1+w_{max}) (1-r_{D^{*}})r_{D^{*}}} \mathcal{F}_1 (w_{max}).
\label{eq:FF-kin-constr}
\eea
In our analysis, we consider the fitted values of the form factor parameters from \cite{Biswas:2021pic}. Lastly, for completion, we would like to list the relations between the BGL form factors and the hadronic form factors \cite{Gambino:2020jvv} :
\bea 
g &=& \frac{2}{m_B + m_{D^*}} V, \\
f &=& (m_B + m_{D^*}) A_1, \\
\mathcal{F}_1 &=& m_B(m_B + m_{D^*})(w-r_{D^*})A_1 - \frac{2 m_B m_{D^*}(w^2 -1)}{1+r_{D^*}}A_2, \\
\mathcal{F}_2 &=& 2 A_0.
\eea 

\begin{table}[t]
\begin{tabular}{|c|c|}
\hline
Form Factor & Type \\
\hline
$n_I$ & $2.6$ \\
$\chi_{1^+}^T(0)$ GeV$^{-2}$ & $ 3.894\times 10^{-4}$ \\
$\tilde{\chi}_{1^-}^T(0)$ GeV$^{-2}$ & $ 5.131\times 10^{-4}$ \\
$\tilde{\chi}_{1^+}^L(0)$  & $ 1.9421\times 10^{-2}$ \\
\hline
\end{tabular}
\caption{Relevant inputs for the outer functions taken from \cite{Bigi:2017jbd}.}
\label{tab:BGL-other-inputs}
\end{table}

\begin{figure}[htp!]
  \centering
  \includegraphics[width=0.45\textwidth]{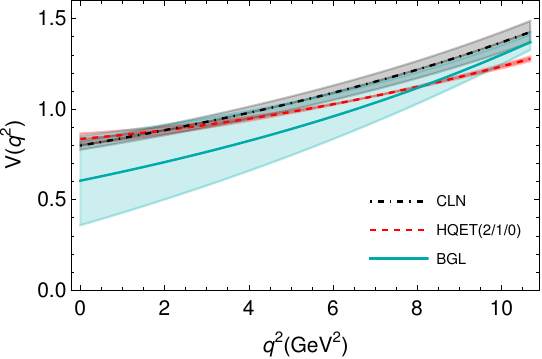}
  \includegraphics[width=0.45\textwidth]{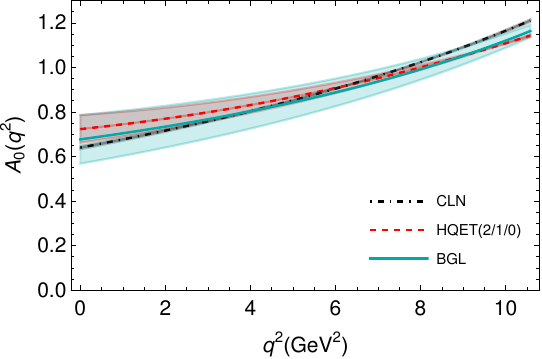}\\
  \includegraphics[width=0.45\textwidth]{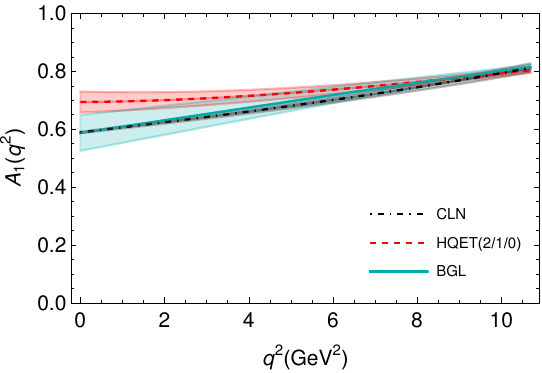}
  \includegraphics[width=0.45\textwidth]{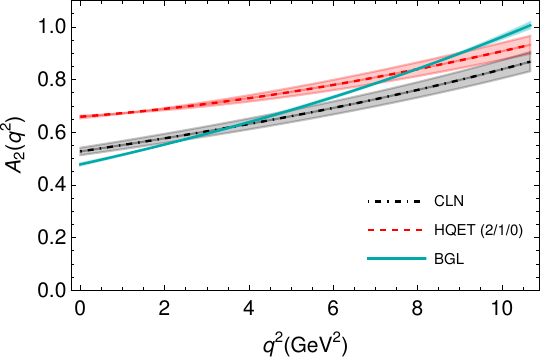}\\
  \includegraphics[width=0.45\textwidth]{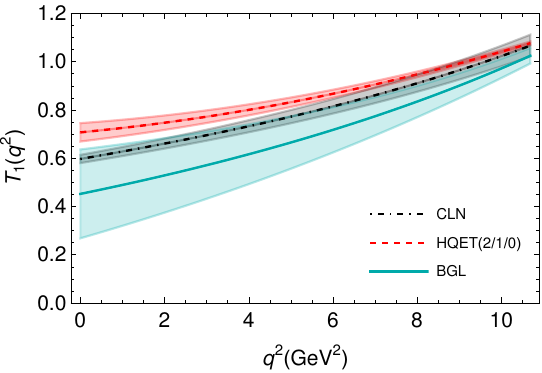}
  \includegraphics[width=0.45\textwidth]{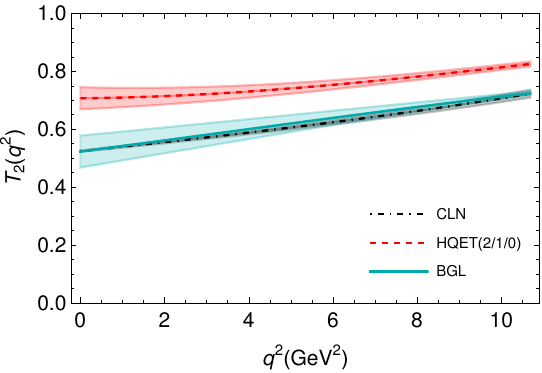}\\
  \includegraphics[width=0.45\textwidth]{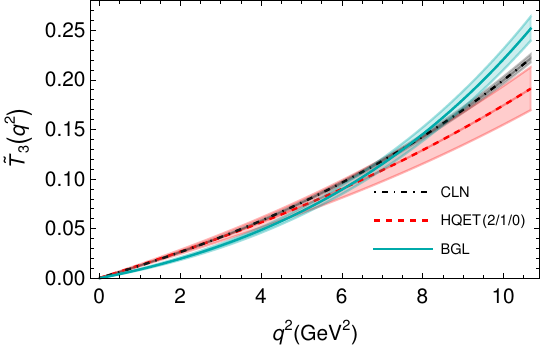}
  \caption{\label{fig:ff}Form factor dependence on $q^2$ for three different form factor parametrizations. The shaded band show the region with the $1\sigma$ upper and lower limits of the form factor parameters listed in Tables~\ref{tab:inputs-others} and \ref{tab:FFparam} are considered without any correlation. For the HQET form factors, we show only the $2/1/0$ scenario following the analysis presented in Ref.~\cite{Iguro:2020cpg}. Here, $\tilde{T}_3 (q^2)$ is defined as $\tilde{T}_3 (q^2) = T_3(q^2) q^2/(m_B^2-{m_D^{*}}^2)$.}
\end{figure}

The form factor dependences on $q^2$ for the various types of parameterizations are shown in Fig.~\ref{fig:ff}.

\bibliography{Nref}
\end{document}